\documentclass[12pt]{article}

\usepackage[margin=1in]{geometry}

\usepackage{listings}
\usepackage{amsthm}
\usepackage{amsmath}
\RequirePackage[numbers]{natbib}
\usepackage[colorlinks,citecolor=blue,urlcolor=blue,filecolor=blue]{hyperref}
\usepackage{graphicx}
 \usepackage{amssymb}
 \usepackage{amsfonts} 
\usepackage{algorithm}
\usepackage{algpseudocode}
\usepackage{colortbl}

\newcommand{\wt}[1]{\widetilde{#1}}

\def\C {\,|\:}

\newcommand\E{\mathbb E}
\renewcommand\d{\mathrm d}

\renewcommand\P{\mathbb P}

\newcommand\R{\mathbb R}

\newcommand\e{\mathrm e}

\newcommand\M{\mathbb M}

\newtheorem{lemma}{Lemma}
\newtheorem{theorem}{Theorem}
\newtheorem{ass}{Assumption}

\theoremstyle{plain} 
\newtheorem{definition}{Definition} 
\newtheorem{example}{Example}
\newtheorem{remark}{Remark}

\def\spacingset#1{\renewcommand{\baselinestretch}%
{#1}\small\normalsize} \spacingset{1}

\begin{document}

\spacingset{1.25}

\def\spacingset#1{\renewcommand{\baselinestretch}%
{#1}\small\normalsize} \spacingset{1}

\date{}
\spacingset{1.25}

\title{\sf Deep Bayes Factors}

\author{Jungeum Kim\footnote{Booth School of Business, University of Chicago, \texttt{jungeum.kim@chicagobooth.edu}}  and Veronika Rockova\footnote{Booth School of Business, University of Chicago, \texttt{veronika.rockova@chicagobooth.edu}}}
 \maketitle
\begin{abstract}
There is no other model  or hypothesis  verification tool in Bayesian statistics that is as widely used as the Bayes factor. We focus on generative models  that are likelihood-free and, therefore, render the computation of Bayes factors (marginal likelihood ratios)  far from obvious. We propose a deep learning estimator of the Bayes factor based on simulated data from two competing models using the likelihood ratio trick. This estimator is devoid of summary statistics and obviates some of the difficulties with ABC model choice. We establish sufficient conditions for consistency of our Deep Bayes Factor estimator as well as its consistency as a model selection tool. We investigate the performance of our estimator on various examples using a wide range of quality metrics related to estimation and model decision accuracy. After training, our deep learning approach enables rapid evaluations of the Bayes factor estimator at any fictional data arriving from either hypothesized model, not just the observed data $Y_0$. This allows us to inspect entire Bayes factor distributions under the two models  and to quantify the relative location of the Bayes factor evaluated at $Y_0$ in light of these distributions. Such tail area evaluations are not possible for Bayes factor estimators tailored to $Y_0$.
We find  the performance of our Deep Bayes Factors competitive with existing MCMC  techniques that require the knowledge of the likelihood function. We also consider variants for posterior or intrinsic Bayes factors estimation. 
We accompany our model selection framework with a model criticism framework based on predictive generative Bayes. We demonstrate the usefulness of our approach on a relatively high-dimensional real data example about determining cognitive biases.
\end{abstract}

\noindent {\it Keywords}: Bayes Factors, Likelihood-free Bayesian Inference, Classification, Implicit Models









\spacingset{1.4}
\section{Introduction}

Bayes factors {(BFs)} are the main staple of Bayesian  hypothesis testing and model comparisons  \citep{jeffreys1935some,berger1985statistical,kass1995bayes}. Defined as  ratios of two marginal   likelihoods, Bayes factors augment the classical likelihood ratio test by reflecting relative predictive performance. In addition, they  quantify the extent to which data warrant prior-to-posterior  change in model odds. 



While Bayes factors have a wide advocacy within the Bayesian community \cite{berger1987testing, savage1972foundations}, there are operational and conceptual connundra. Besides the difficulty in choosing plausible default proper priors for model comparisons \citep{o1995fractional, berger1996intrinsic}, Bayes factor computation itself can  be daunting. Practitioners may resort to convenient approximations such as the BIC \citep{schwarz1978estimating} which, unfortunately, may not  easily lend itself to an analysis that seeks to take advantage of  background knowledge through informative priors.  Indeed, the BIC approximates the ``default" Bayes factor that is based on a unit-information prior \citep{kass1995reference}.  When likelihoods are, in fact, available, various analytical approximations \citep{jeffreys1998theory, kass1995bayes,raftery1996approximate} or simulation approaches (including  Bridge sampling and variants thereof  \cite{meng2002warp, gelman1998simulating}) are available. Our methodology targets situations when these approaches cannot be readily applied. 
We focus  on the feasibility of  Bayes factor computation  in the context of simulation-based (generative) models. {Bayes factors are used broadly across various disciplines \citep{geweke1999using,bridges2009Bayesian,minin2003performance,lee2014Bayesian} including machine learning where the use of marginal likelihoods for model comparisons is increasingly more appreciated   \citep{qi2001investigation,immer2021scalable,dutordoir2020Bayesian}.} Our purpose is to broaden the reach of Bayes factors to contexts involving  complex models where marginal likelihood evaluations would seem nonviable.



A wide range of sampling-based Bayes factor estimation techniques exists, largely when the likelihood is accessible. Rubin \citep{rubin1987comment,db1988using} emphasized importance sampling for marginal likelihoods, noting however its potential instabilities  when the posterior distribution is sharply peaked relative to the prior. Savage-Dickey representations, as laid out by \cite{verdinelli1995computing}, cater specifically to nested models. Additional methods worth noting include harmonic means, Laplace approximation, stochastic substitution, nested sampling, and reverse logistic regression \citep{raftery2006estimating, tierney1986accurate,gelfand1990sampling,geyer1994estimating, chopin2010properties}. Comparative insights into these methodologies can be sourced from \cite{marin2007bayesian}.
Our contribution  centers primarily on situations when likelihoods are implicit  (analytically intractable), not only when the computation of marginal likelihoods presents a challenge.  A crucial requirement for our methodology is that the  likelihood can be simulated from. Bayesian inference in such likelihood-free scenarios has been accessed through the utilization of Approximate Bayesian Computation (ABC) \citep{tavare1997inferring, pritchard1999population, beaumont2002approximate}. A default ABC approach to estimating Bayes factors involves calculating the ratio of  acceptance rates under considered models using ABC reference tables \citep{grelaud2009abc,toni2010simulation,didelot2011likelihood}. However, \cite{robert2011lack} show that ABC for Bayes factor calculations can be biased {when summary statistics are insufficient for model selection}. We investigate the usefulness of deep learning for Bayes factor estimation. The core idea is fairly simple. We deploy  the likelihood-ratio trick from machine learning
\citep{hastie2009elements,sugiyama2012density,mohamed2016learning}  to estimate marginal likelihood ratios via binary classification. The effectiveness of  classification-based density ratio estimates has been appreciated  in the Bayesian context for Metropolis-Hastings simulation \citep{kaji2022metropolis}, posterior estimation \citep{thomas2022likelihood}, and ABC \citep{wang2022approximate}. We were intrigued by the possibility of leveraging contrastive learning for Bayes factor computation in models where prior and likelihood simulations are possible.  

{In this work, we introduce Deep Bayes Factor {(DeepBF)}, a neural classifier trained on simulated datasets to learn a mapping whose functional constitutes a Bayes factor estimator.} {The trained} mapping enables Bayes factor evaluations  over the {\em entire data domain} (not just observed data)  with negligible additional computational cost. This  feature allows us to compute distributions of Bayes factors under each hypothesis and perform model selection {\em inference} using tail area evidence measures. Many Bayes factor computation methods are tailored to the observed data \citep{meng1996simulating, meng2002warp}  which allow only point-evaluations at observed data as opposed to inference using plausible data under each hypothesis. {We support our development theoretically by providing sufficient conditions under which our Deep Bayes Factors are consistent estimators of Bayes factors as well as sufficient consistency conditions for model selection.} Viewing the Bayes factor through the lens of binary classification aligns with \cite{pudlo2016reliable}, who recast ABC  model selection as a classification problem. They employ random forests to select a model by a majority vote. Instead, we focus on binary classification where the purpose is to learn marginal likelihood ratios. Contrary to the method in \cite{pudlo2016reliable}, our strategy circumvents a secondary learning phase for gauging model posterior estimates, delivering results in only one stage. There have several attempts to apply deep learning to the problem of Bayes factor estimation from other perspectives. {For example, \cite{jia2020normalizing} used normalizing flow approach to estimate the marginal likelihood, and \cite{xing2022improving} suggested to use f-GAN to improve bridge estimators \citep{ali1966general}. While their approaches rely on MCMC sampling from the posteriors to obtain training datasets, our DeepBF is trained solely on simulated data from two competing models, not the posteriors.}

{Our second contribution is furnishing our methodology with a toolkit  for evaluating goodness of model fit as well as goodness of Bayes factors for model selection inference. Our exhaustive evaluation toolkit provides a more comprehensive framework for informed model selection and enables effective objective comparisons of various Bayes factor estimators. Firstly, since model selection cannot be entirely separated from model criticism, we discuss several goodness-of-fit strategies using contrastive learning, which can be used as a prerequisite to Bayes factor evaluations. In particular, we develop a contrastive learning approach to model adequacy assessment based on comparing observed data with data generated from a posterior predictive distribution. We develop a score based on an estimated classifier and gauge model adequacy based its the departure away from $1/2$. While Bayes factor comparisons indicate which of the two considered models is more plausible, our model adequacy framework helps determine whether considered models   themselves are
compatible with the observed data.
Secondly, we evaluate goodness of Bayes factors with visualization tools and validation measures including ROC curves and estimated model prior quality measures. Additionally, while the magnitude of Bayes factors evaluated at observed data traditionally indicates the strength of model evidence \citep{jeffreys1998theory,kass1995bayes,lee2014Bayesian}, recent works complement this by quantifying the amount of evidence using a tail probability \citep{baskurt2013hypothesis} derived from simulated data \citep{garcia2005calibrating, gu2016error, schonbrodt2018bayes, schad2022workflow}. This improved understanding can lead to calibrations of Bayes factor decision thresholds. Our validation measures, therefore, include the distributional quality of Bayes factor estimates that does not penalize systematic bias or imprecise scale such as area under ROC curve (AUC) and MSE of surprise measures. These evaluations are further enhanced by our rapid and simultaneous computation of our DeepBF estimator.} We demonstrate performance of Deep Bayes Factors on several well-studied examples where ABC model choice faces challenges. Lastly, we apply our approach to a real data example on testing for the presence of cognitive biases.
 
 The paper is structured as follows. In Section \ref{sec:method}, we review ABC in the context of Bayes factor estimation and introduce our Deep Bayes Factor estimator. Theoretical justifications for Deep Bayes Factors are presented in Section \ref{sec:theory}. Section \ref{sec:gAI} discusses a model-check framework for generative models.
Section \ref{sec:numerical} {presents the evaluation toolkit for goodness of Bayes factors. Section \ref{sec:sim} and \ref{sec:real2}} describe an application of our method to both simulation and real datasets.
The paper concludes in Section \ref{sec:conclusion}.

\paragraph{Notation}
For a distribution $P$ with a density $p$, we denote with   $P^\infty$ the  infinite product measure. For two densities $p^*$ and $p$, we define $K(p^*, p)=\int p^*\log(p^*/p)$ and $V(p^*,p)=\int p^*(\log(p^*/p))^2$.
 With $P_1$ and $P_2$ we denote the measures with densities $\pi(\cdot \C M_i)$ under the model $M_1$ and $M_2$. With $E_1$ and $E_2$ we denote the corresponding expectations. For two discriminators $D_1$ and $D_2$, we define a Hellinger-type discrepancy $d_H(D_1,D_2)=\sqrt{h(D_1,D_2)^2+h(1-D_1,1-D_2)}$ where 
$h(D_1,D_2)=\sqrt{(E_1+E_2)(\sqrt{D_1}-\sqrt{D_2})^2}$. 

\section{Approximating Bayes Factors via Classification}\label{sec:method}
Our setup consists of an observed  data vector $Y^{(n)}_0=(Y_1^0,\dots, Y_n^0)\in\mathcal Y^{n}$ which could have plausibly arrived from a model $M_1$ or a model $M_2$.
These two models are specified through conditional densities  $\pi_1(\cdot \C\theta_1)$ and $\pi_2(\cdot \C\theta_2)$ with unknown  model parameters $\theta_1\in\Theta_1\subseteq\R^{d_1}$ and  $\theta_2\in\Theta_2\subseteq\R^{d_2}$ that are 
assigned  prior  distributions $\pi_1(\theta_1)$ and $\pi_2(\theta_2)$, respectively. The Bayes Factor  is then defined as a ratio of the marginal likelihoods evaluated at $Y^{(n)}_0$
\begin{equation}\label{eq:bf}
BF_{1,2}(Y^{(n)}_0)=\frac{\int \pi_1(Y^{(n)}_0\C\theta_1)\pi_1(\theta_1)\d\theta_1 }{\int \pi_2(Y^{(n)}_0\C\theta_2)\pi_2(\theta_2)\d\theta_2}=\frac{\pi(Y^{(n)}_0\C M_1)}{\pi(Y^{(n)}_0\C M_2)}.  
\end{equation}

 When likelihoods  are analytically unavailable, but can be simulated from, a popular approach to estimating Bayes factors  is ABC 
 \citep{grelaud2009abc,toni2010simulation,didelot2011likelihood,robert2011lack}.
We review this approach to build a foundation for our own development.

\subsection{ABC Model Choice}
ABC was originally introduced in population genetics \citep{tavare1997inferring, pritchard1999population, beaumont2002approximate} for inference about evolution of species. Instead of sampling directly from the posterior, ABC draws candidate parameters from the prior and accepts them if  simulated data from the likelihood (given candidate parameters) closely aligns with the observed data. For a comprehensive introduction to ABC  we refer to \cite{sunnaaker2013approximate}. With some caution, ABC can be also used for model choice.

 A common ABC approach to Bayes factor estimation  \citep{robert2011lack}  is calculating the ratio of acceptance rates for the two models using ABC reference tables involving simulated  data from the marginal likelihood. During the $i^{th}$ of $N$  total  ABC iterations, one samples the model index $k^{(i)}\in\{1,2\}$ based on prior model probabilities $\pi(M_1)$ and $\pi(M_2)$. Second,  parameters $\theta^{(i)}$ are drawn from the prior $\pi_{k^{(i)}}(\cdot)$   and, finally, 
 an $(n\times 1)$ fake data vector $Y^{(i)}$ is simulated from  $\pi_{k^{(i)}}(\cdot \C\theta^{(i)})$.  The estimate of the Bayes factor based on the ABC lookup table  $(k^{(i)},\theta^{(i)},Y^{(i)})_{i=1}^N$ is then obtained as 
 \begin{equation}\label{eq:abc1}
\widehat{BF}^{\rm ABC}_{1,2}(Y_0^{(n)}) = \frac{\pi(M_2) \sum_{i: k^{(i)}=1}^N \mathbb I\{\rho[\eta(Y^{(i)}),\eta(Y_0^{(n)})]\leq \epsilon\} }{\pi(M_1) \sum_{i: k^{(i)}=2}^N
 \mathbb I\{\rho[\eta(Y^{(i)}),\eta(Y_0^{(n)})]\leq \epsilon\}},
\end{equation}
 where $\eta(\cdot)$ is a summary statistic used for ABC, $\rho$ is a distance criterion, and $\epsilon>0$ is a threshold.

The implementation of $\widehat{BF}^{\rm ABC}_{1,2}(Y_0^{(n)})$ in \eqref{eq:abc1} is straightforward and bypasses  direct calculations of the likelihood which makes it applicable to a wide range of complex and non-standard models. For Gibbs random fields, \cite{grelaud2009abc} demonstrated  convergence of \eqref{eq:abc1} to the true Bayes factor.
  \cite{robert2011lack} cautioned against inherent bias in ABC-based Bayes factor estimates due to the information loss induced by summary statistics. Their core intuition is that, even if the summary statistic $\eta(\cdot)$ is sufficient for both $\pi_1(\cdot \C \theta_1)$ and $\pi_2(\cdot\C\theta_2)$, it may not be sufficient for $\{(k, \pi_k(\cdot\C\theta_k))\}_{k=1,2}$, rendering it unsuitable for model selection.
 

Previous studies in ABC literature have also highlighted the relationship between Bayes factor estimation and the classification problem. Given a set of  \( K \) competing models $M_1,\cdots, M_K$, \cite{fagundes2007statistical} estimated the posterior probability  $ \pi(M_k\C Y_0^{(n)}) $ of each model using multinomial logistic regression trained on an ABC filtered reference table using the model category $k^{(i)}$ as a label and summarized data $\eta(Y^{(i)})$  as predictors. A similar concept is presented in \cite{pudlo2016reliable} who employed two distinct random forests in two stages. First, a random forest classifier is trained. In the second stage, a separate random forest regression is deployed, targeting the error rate of the initial classifier as an estimator of \( \pi(M_k\C Y^{(n)}_0) \). Our method operates in a single stage, where a more flexible classifier is deployed and regarded as a function of the Bayes factor. While we also frame Bayes factor estimation as classification, we do not rely on summary statistics and acceptance thresholds. 

\subsection{Deep Bayes Factors}\label{eq:our_method}

The core idea behind  Deep Bayes Factors is fairly simple and rests on  the likelihood-ratio trick. To learn density ratios, this trick employs binary classification on two contrasting datasets, which are equipped with a binary label \citep{hastie2009elements,sugiyama2012density,mohamed2016learning}. The likelihood-ratio trick has been widely utilized in various contexts including generative adversarial networks \citep{goodfellow2020generative}, noise-contrastive estimation \citep{gutmann2010noise}, and mutual information estimation \citep{belghazi2018mutual}, among others. In the Bayesian literature, \cite{kaji2022metropolis} used the likelihood-ratio trick for conditional likelihood ratio estimation inside the Metropolis-Hastings algorithm for posterior simulation. Here, we consider marginal likelihood ratio estimation for model comparisons. 

Throughout this section we denote the $(n\times 1)$ random data vector   simply as $Y$.
Denote by $\mathbb D$ a set of all classification functions $D:\mathcal{Y}^{n}\rightarrow (0,1)$. 
We regard discerning   $M_1$ from $M_2$ as a binary classification problem with labels $1$ (model $M_1$) and $0$ (model $M_2$) with the optimal solution
\begin{equation}\label{eq:D}
D^*=\arg\max_{D\in\mathbb D} [E_{Y\sim P_1}\log D(Y) +E_{Y\sim P_2}\log (1-D(Y))].
\end{equation}
Recall that $P_1$ and $P_2$ correspond to the measures with marginal likelihoods $\pi(\cdot\C M_1)$ and $\pi(\cdot\C M_2)$, respectively.
The optimal solution $D^*$ satisfies (Proposition 1 in \citep{goodfellow2020generative})
\begin{equation}\label{eq:optimal_D}
\frac{D^*(Y)}{1-D^*(Y)}=\frac{\pi(Y\C M_1)}{\pi(Y\C M_2)}=BF_{1,2}(Y)\quad\text{for any $Y\in{\rm Supp}(P)$},
\end{equation}
where ${\rm Supp}(P)$ is the support of the mixture $P=\pi(M_1)P_1+\pi(M_2)P_2$.
Our goal is to approximate $D^*$ by considering a {\em rich enough} class of classifiers $\mathcal D_n\subset\mathbb D$ (e.g. deep learning classifiers; see Remark \ref{rem:training} and Section \ref{sec:comp}) and by replacing the expectations in \eqref{eq:D} with empirical averages over large enough simulated data sets. Given the training sample size $T$ for each model, suppose that we simulate $Y_1,\dots, Y_T\in \mathcal{Y}^{n}$ from model $M_1$, labeling them as $1$, and $\tilde Y_1,\dots, \tilde Y_T\in\mathcal{Y}^{n}$ from model $M_2$, labeling them as $0$. On these simulation sets, {we define 
\begin{equation}\label{eq:estimator}
\hat D_T=\arg\max_{D\in\mathcal D_n} \M_T(D)
\end{equation}
as the classification function in $\mathcal D_n$ that maximizes $\M_T(D)$  defined as }
\begin{equation}\label{eq:loss}
\M_T(D)=  \P_{1,T} \log D + \P_{2,T} \log (1-D),
\end{equation}
where $\P_{1,T} f = \frac{1}{T}\sum_{i=1}^T f(Y_i)$ and $\P_{2,T} f = \frac{1}{T}\sum_{i=1}^T f(\tilde Y_i)$. {Finding $\hat D_T$ is} equivalent to learning a binary classifier (such as logistic regression) using $Y_i$ and $\tilde Y_i$ as predictors. The optimal solution $\hat D_T$  of \eqref{eq:loss} will be an approximation of $D^*$ in \eqref{eq:D}. Our estimation strategy proceeds in two steps: (1) learn $\hat D_T$ by solving \eqref{eq:estimator} using simulated data, (2) plug-in the observed data vector $Y_0^{(n)}$ into a simple transformation of $\hat D_T$. More precisely, our estimator of the Bayes factor can be obtained by replacing $D^*$ in \eqref{eq:optimal_D} with $\hat D_T$  as
\begin{equation}\label{eq:est}
\widehat{BF}_{1,2}(Y_0^{(n)})=\frac{\hat D_T(Y_0^{(n)})}{1-\hat D_T(Y_0^{(n)})}.
\end{equation}
One of the strengths of our approach is  that the estimation is {\em not restricted to} $Y_0^{(n)}$. By applying the same $\hat D_T$   in \eqref{eq:est} on any $Y\in{\rm Supp}(P)$, we can get an estimator of $\widehat{BF}_{1,2}(Y)$.

\begin{figure}[!t]
\centering
\begin{minipage}{\linewidth}
\begin{algorithm}[H]
\caption{{Deep Bayes Factor  Estimator }}\label{alg:original}
\centering
\spacingset{1.1}
\small
\resizebox{\linewidth}{!}{
\begin{tabular}{l l}
\multicolumn{2}{l}{\textbf{Fixed Input}: The observed data $Y_0^{(n)}\in \mathcal{Y}^{n}$, training sample size $T$, number of iterations $M$,}\\
\multicolumn{2}{l}{\hspace{2.2cm}$s = T/M$ (mini-batch size)}\\
\multicolumn{2}{l}{\textbf{Learnable Input}: Randomly initialized classifier function $D^{(0)}$ parametrized by $\phi^{(0)}$}\\
\multicolumn{2}{c}{ \cellcolor[gray]{0.9}{\bf Training}  }\\
\multicolumn{2}{l}{For $t=1,...,M$}\\
\multicolumn{2}{l}{\qquad Sample $\theta_{1,j}$ from $\pi_1(\cdot)$ and $Y_j$ from $\pi_1(\cdot\C\theta_{1,j})$ for $j=1,...,s.~~~~~~~~~~~~~~~~~~~~~~~$}\\
\multicolumn{2}{l}{\qquad Sample $\theta_{2,j}$ from $\pi_2(\cdot)$ and $\tilde Y_j$ from $\pi_2(\cdot\C\theta_{2,j})$ for $j=1,...,s$.}\\
\multicolumn{2}{l}{\qquad Compute $\mathbb{M}_s(D^{(t-1)}) = [\sum_{j=1}^s\log D^{(t-1)}(Y_j) +\log (1-D^{(t-1)})(\tilde Y_j)]/s$}  \\
\multicolumn{2}{l}{\qquad Update $D^{(t)}$ {by updating $\phi^{(t)}$ by using $\nabla_\phi \mathbb{M}_s(D^{(t-1)})$ via} a stochastic optimizer such as Adam}\\
\multicolumn{2}{c}{ \cellcolor[gray]{0.9}{\bf Bayes factor Estimation on the original data $Y_0^{(n)}$}  }\\
\multicolumn{2}{l}{Set $\hat D_T= D^{(M)}$}\\
\multicolumn{2}{l}{Return  $\widehat{BF}_{1,2}(Y_0^{(n)}) = {\hat D_T(Y_0^{(n)})}/(1-\hat D_T(Y_0^{(n)}))$}
\end{tabular}}
\end{algorithm}
\end{minipage}
\end{figure}

\begin{remark}  
{Marginal likelihood ratio estimation is intimately linked to the task of  estimating the \(f\)-divergence 
$d_f(P_1|P_2) = \int \pi(\cdot\C M_1) f\left(\frac{\pi(\cdot\C M_2)}{\pi(\cdot\C M_2)}\right), $ where \(f\) is a convex function satisfying \(f(1) = 0\). A prevalent approach for evaluating the divergence is through the Fenchel conjugate \(f^*\), which yields a dual formula $d_f(P_1|P_2) \geq  \sup_{ F\in\mathbb{T}} E_1[F ]-E_2[f^*(F )] $ {  for a rich class functions $\mathbb{T}$.} For instance, the original GAN \cite{goodfellow2020generative} was extended to the f-GAN framework \citep{nowozin2016f} by considering the \(f\)-distance between the original and fake distributions and optimizing over \(\mathbb{T}\) using deep neural networks. It is known that choosing a suitable $f$ for specific tasks can lead to more stable optimization \citep{shannon2020properties}. Because equality is achieved when \(F^*  = f'\left(\frac{p_1 }{p_2 }\right)\), inverting \(f'\) in \(F^*\) could yield the desired likelihood ratio. We adopted  the classification formulation   \eqref{eq:D}, acknowledging that other estimators could be obtained by choosing  a suitable \(f\) and inverting \(f'\). For a more comprehensive overview of likelihood ratio estimation through f-divergences, we refer the reader to \cite{sugiyama2012density}.} 
\end{remark}

In our demonstrations, we include all  simulated samples $Y_t$ and $\tilde Y_t$ for $1\leq t\leq T$ as our training data for $\hat D_T$. Note that we do not require independence of the entries in $Y$,
rendering this approach suitable for time series or otherwise correlated data.
We could tailor the estimator to  $Y_0^{(n)}$ by training the classifier on a screened-out ABC lookup table or using  reinforcement-learning \citep{sean}.  
\begin{remark}\label{rem:training}
Throughout this paper, we primarily focus on deep learning classifiers $\mathcal D_n=\{d_\phi(\cdot):\phi\in\R^d\}$ parametrized by $\phi$. We describe the architecture choice in Section \ref{sec:comp} and provide the code in Section \ref{section:network}.
Utilizing stochastic optimization, we construct the training dataset on-the-fly. To train with as many samples as possible in a memory-efficient manner, we generate a new mini-batch of size \(T/M\) from each model during every iteration, where \(M\) represents the total number of iterations.   A similar idea was used by \cite{shin2024generative}, to avoid over-fitting since generating a new (mini)-batch for every single iteration maximizes the number of simulated data points used for the training. The described algorithm is in Algorithm \ref{alg:original}. {This algorithm produces a trained network $D^{(M)}$ after $M$ mini-batch iterations that we substitute for the estimated classifier $\hat D_T$.} 
Algorithm \ref{alg:original} shows Bayes factor estimation evaluated on a single dataset $Y_0^{(n)}$. Note that after $\hat D_T$ has been trained, the Bayes factor estimator can be evaluated {\em on any  dataset} $Y$. 
\end{remark}

\begin{example}[Toy Example] \label{ex:toy} We apply the Deep Bayes Factor estimator \eqref{eq:est} to three toy examples in an iid data setup, for which closed-form Bayes factors exist: Data 1) a comparison between Negative Binomial and Poisson distributions \citep{BERN},  Data 2) Gaussian mixture data with a Gaussian prior on the means, and Data 3) a nested model test distinguishing between hierarchical Bayesian models. More details are in Section \ref{example_details}. The results are visualized in Figure \ref{fig:two_dim_good}, which shows high similarity between the true Bayes factors and the estimated ones on a large portion of the data domain that contains most of the support for both models. For the actual demonstration of the samples, see Figure \ref{fig:two_dim_good2} in Section \ref{sec:additional_plots}.
\end{example}

{\spacingset{1.5}
\begin{figure}[!t]
    \centering
    \includegraphics[width=0.7\linewidth]{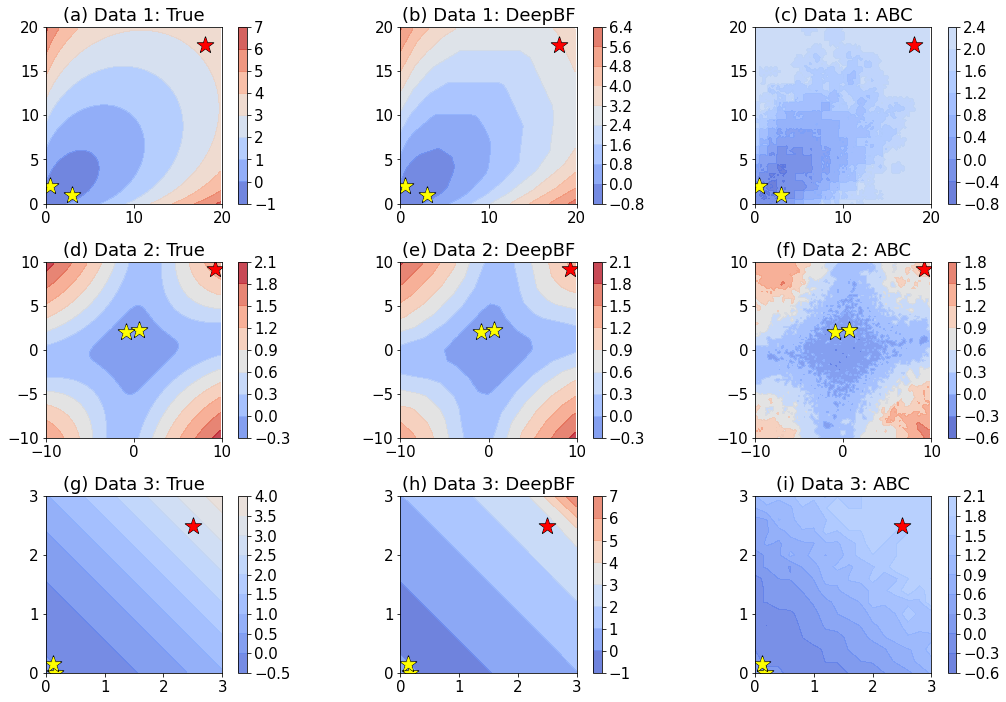}
    \vspace{-.5cm}
    \caption{True and estimated {BF's} on the $\log_{10}$ scale ({DeepBF} and ABC) evaluated on a fine  grid of values $Y=(Y_1,Y_2)'$.  First row: Negative Binomial vs Poisson distributions. Second row: {Gaussian vs Gaussian mixture priors}. Third row: Hierarchical  exponential distribution.  {Stars indicate outliers (red) and inliers (yellow) used for the model validation demonstration (see Example \ref{ex:toy3}).}}
    \label{fig:two_dim_good}
\end{figure}
}


\subsection{Deep Bayes Factor Extensions}\label{sec:variants}
Deep Bayes Factors can be tailored to alternative Bayes factor formulations that leverage data splitting. 
 Decomposing the data vector as \( Y = (X, Z)' \), where \( X \in \mathcal{Y}^{n_x} \) and \( Z \in \mathcal{Y}^{n_z} \), recall the conditional Bayes factor  for $Z$ given $X$ (aka the partial Bayes factor PBF \citep{lempers, o1995fractional, berger1996intrinsic}) 
\begin{equation}\label{eq:deepCond}
  PBF_{1,2}(Z|X) = \frac{\pi(Z|X,M_1)}{\pi(Z|X,M_2)}= {BF_{1,2}(Y)}{BF_{2,1}(X)},
\end{equation}
where the last equality is due to $\pi(Z|X,M_i) \pi(X|M_i)= \pi(Z,X|M_i)$. PBF's were originally designed to handle non-informative (improper) priors or prior sensitivity by using the posterior $\pi_k(\theta_k|X)$ instead of the prior $\pi_k(\theta_k)$ for $k=1,2$. 
The characterization \eqref{eq:deepCond} allows us to estimate PBF's by deploying two Deep Bayes Factors ({DeepBF's}).  Another Bayes factor variant is the posterior Bayes factor, defined as 
 $PBF_{1,2}(Z|X)$ in \eqref{eq:deepCond} where  $X=Y$ and $Z=Y$. While one can estimate the posterior Bayes factor using deep learning via \eqref{eq:deepCond}, it may not be preferred over alternatives because of its double use of the data. Another way of  handling improper priors that is robust with respect to the choice of one particular split $(X,Z)$ 
is the intrinsic Bayes factor (IBF, \cite{berger1996intrinsic})  which reflects all   $K=\binom{n}{n_x}$  possible data subsets $Y(i)\in \mathcal{Y}^{n_x}$ of size $n_x$ and aggregates them through a geometric or arithmetic average. Seen as an actual Bayes factors for an \emph{intrinsic prior} \cite{berger1996intrinsic}, the arithmetic IBF is defined as {
\begin{equation}\label{eq:ibf}
    ABF^{n_x}_{1,2}(Y) = \left(\sum_{i=1}^{K}PBF_{2,1} (Y_{\backslash i}|Y(i))\right)/K=\left(\sum_{i=1}^{K}{B_{1,2}(Y)}{B_{2,1}(Y(i))}\right)/K,
\end{equation}
where $Y_{\backslash i}=Y\backslash Y(i)$ and $Y(i)\in \mathcal{Y}^{n_x}$ is $i$-th   out of $K=\binom{n}{n_x}$  possible subsets of size $n_x$. 
{GBF is defined similarly, using a geometric average instead.}  Because our Deep Bayes Factor estimator for $BF_{2,1}(X)$ can be evaluated for {\em any dataset} $X$ of size $n_x$,  a single deep learning mapping   is sufficient to compute $BF_{2,1} (Y(i))$ for {\em all} $i=1,...,K$. Therefore, by training such a network in addition to the original Bayes factor estimator of $BF_{1,2}(Y)$ (as  outlined in Section \ref{eq:our_method}), one can estimate both ABF's and GBF's by simply plugging-in different subsets of data into the same trained mapping $\widehat {BF}_{2,1}(\cdot)$. 
{\spacingset{1.5}
\begin{figure}
    \centering
    \includegraphics[width=\linewidth]{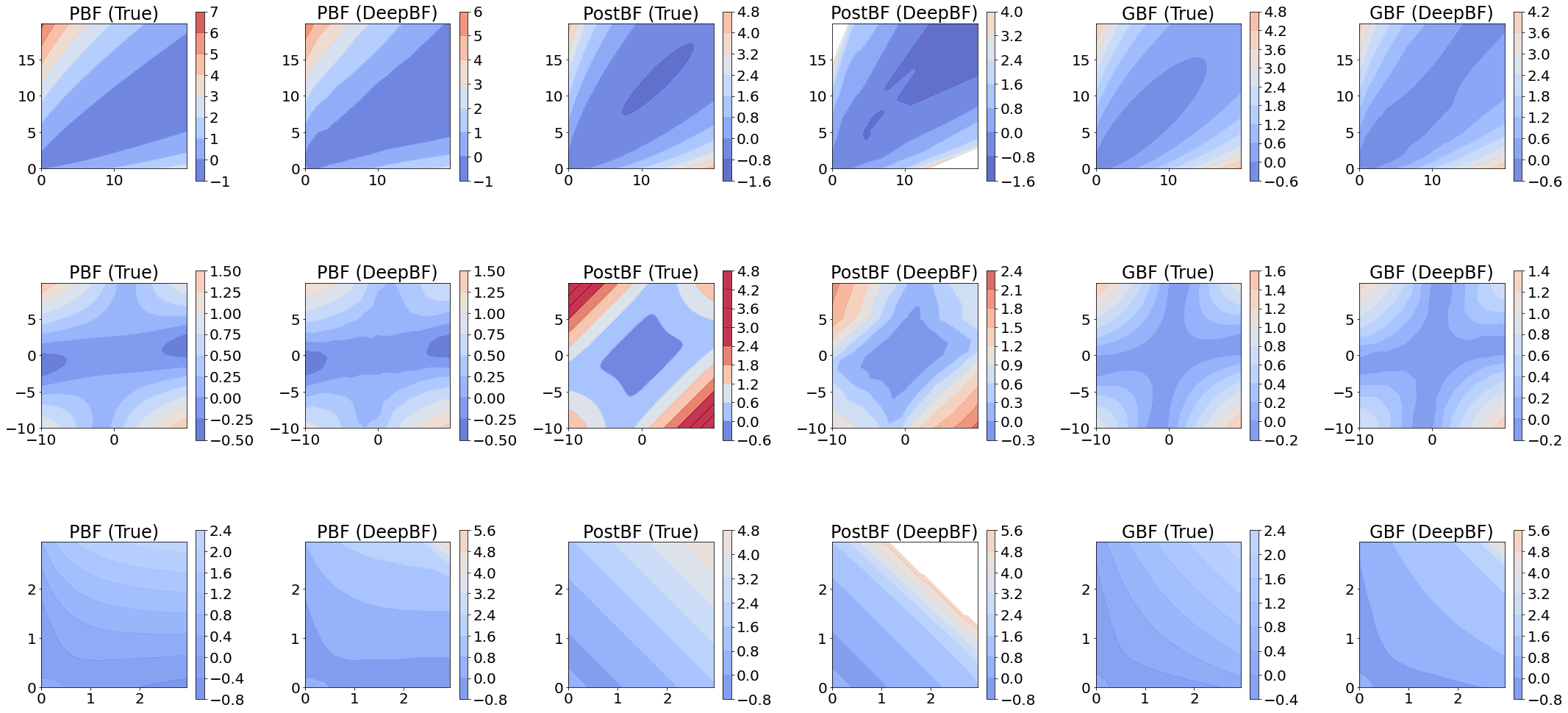}
    \vspace{-1cm}
    \caption{Examples of {Bayes factor variants}  ($\log_{10}$ scale): PBF is a partial BF, PostBF is a posterior BF and GBF is a geometric variant of the intrinsic BF. The blanc area corresponds to numerical overflow (as elaborated on in Figure \ref{fig:robust} and \ref{fig:two_dim_good2} (Section \ref{sec:additional_plots}).  The three datasets are from Example \ref{ex:toy} with $n=2$. }
    \label{fig:partial1}
\end{figure}
}
 \begin{example}[Toy Example] \label{ex:toy_cont} We continue our toy example exploration  started in Example \ref{ex:toy}. As a preliminary investigation of the usefulness of our techniques to estimate various Bayes factors, we visualize the result of each method in Figure \ref{fig:partial1} for all three datasets from Example \ref{ex:toy} for $n=2$ (for wider range, see Figure \ref{fig:robust} in Section \ref{sec:additional_plots}). The scatter-plots of   true and estimated Bayes factors based on the simulated datasets are in Figure \ref{fig:partial3} (Section \ref{sec:additional_plots}). Overall, both figures show  similarity between the estimated and true Bayes factor variants. For Data 3, our technique tends to overestimate, especially for $Y\sim M_1$. While some discrepancies are observed in some areas in Figure \ref{fig:partial1}, datasets corresponding to those areas are less likely to be generated.\footnote{For example, out of 3,000 randomly generated samples, the estimates on 0.35\% of the samples were machine infinity for PBF, and 0.9\% and 0.23\% for PostBF and GBF respectively.} In both Figures GBF demonstrates most stable results. Although not shown in Figure \ref{fig:partial1}, ABF produces similar outputs. In subsequent analyses, we found numerical instability may arise if both \(BF_{1,2}(Y)\) and \(BF_{2,1}(X)\) in   \eqref{eq:deepCond} yield values that are either extremely large or near zero.  
\end{example}

\section{Deep Bayes Factor Consistency}\label{sec:theory}
Traditionally, consistency of Bayes factors refers to the simultaneous occurrence of the two stochastic convergence phenomena  implying an overwhelming amount of evidence for/against one of the two considered models as $n\rightarrow\infty$.  

While the methodology in Section \ref{eq:our_method} does not necessarily assume iid observations, our theory  does.
For observations $Y_0^{(n)}=(Y_1^0,\dots, Y_n^0)'$ arriving as iid realizations from $P^*$ with a density $p^*$, our goal is to distinguish between two composite hypotheses
\begin{equation}
H_0:\,\, p^*\in \mathcal F_0\quad\text{against}\quad H_1:\,\, p^*\in \mathcal F_1,\label{eq:test}
\end{equation}
where $\mathcal F_0=\{\pi_1(\cdot\C\theta_1):\theta_1\in\Theta_1\}$ and $\mathcal F_1=\{\pi_2(\cdot\C\theta_2):\theta_2\in\Theta_2\}$.
The likelihoods under each model are defined implicitly but are parametrized by finite-dimensional parameters. Below, we define the rate of Bayes factor divergence for the testing problem \eqref{eq:test}.
\begin{definition}(Bayes Factor Convergence  Rate)\label{def:consist}
For the testing problem \eqref{eq:test}, we say that the Bayes factor $BF_{1,2}(Y_0^{(n)})$ is consistent at a rate $1/\nu_n$ if 
$$
\lim_{n\rightarrow\infty} P^\infty_* (\log BF_{1,2}(Y_0^{(n)}) <1/\nu_n)=0\quad\text{when $p^*\in\mathcal F_0$} 
$$
and
$$
\lim_{n\rightarrow\infty} P^\infty_* (\log BF_{1,2}(Y_0^{(n)}) >-1/\nu_n)=0 \quad\text{when $p^*\in\mathcal F_1$} 
$$
for a sequence $\nu_n\rightarrow 0$.
\end{definition}
Definition \ref{def:consist} extends the classical notion of Bayes factor consistency (see e.g Section 2 of \cite{mcvinish2009bayesian}) by quantifying the speed of divergence to $\pm\infty$.
There exists a variety of results on Bayes factor consistency in non-parametric (density estimation and/or iid) problems that are based on verifying sufficient conditions related to posterior consistency and posterior convergence rates (see \cite{verdinelli1998bayesian,   walker2004priors, ghosal2008nonparametric, mcvinish2009bayesian, tokdar2010bayesian}).   One of these sufficient conditions, the Kullback-Leibler (KL) property (under which the prior puts positive mass on all Kullback-Leibler neighborhoods of the true value of parameter) is  fundamental to Bayes factor consistency. We consider a similar assumption.
\begin{ass} \label{ass:KL}
Define   $p^{(1)}_{\theta_1}= \pi_1(\cdot\C  \theta_1)$ and $p^{(2)}_{\theta_2}= \pi_2(\cdot\C \theta_2)$ and, for some $c>0$,  
$$
K_n^{(1)}=\left\{\theta_1\in\R^{d_1}:K(p^*, p^{(1)}_{\theta_1} )<c\,n^{-1},V(p^*, p^{(1)}_{\theta_1} )<c\,n^{-1}\right\}.
$$
Similarly, define $K_n^{(2)}$ as  the above with  $p^{(1)}_{\theta_1}$ replaced by $p^{(2)}_{\theta_2}$. Assume that  for some $C>0$ we have (a) $\pi_1\left(K_n^{(1)}\right)\geq Cn^{-d_1/2}$ when 
$p^*\in\mathcal F_0$  and (b)
$\pi_2\left(K_n^{(2)}\right)\geq Cn^{-d_2/2}$ when $p^*\in\mathcal F_1$. 
\end{ass}
Assumption \ref{ass:KL} essentially states that the prior  assigns enough mass around a shrinking neighborhood of $p^*$ under both considered models. \cite{walker2004priors}
characterized the Kullback-Leibler property as assigning a positive mass to a sufficiently large KL neighborhood and shows that  Bayes factor always supports the model with a smaller KL gap, not necessarily requiring that the model arrived from either of the two models. 

Before we establish  model selection consistency of the Deep Bayes Factor estimator \eqref{eq:est}, we first focus on its consistency as a Bayes factor estimator. Theoretical results exist for classifiers obtained by minimizing the entropy loss \citep{kaji2022metropolis, kaji2020adversarial}. The technique in \cite{kaji2022metropolis} relies on 
properties of certain Hellinger-type discrepancy measures for classifiers. We can write the true Bayes factor as a functional of the ``true" discriminator $D^*$ through the relationship $BF_{1,2}=D^*/(1-D^*)$ in \eqref{eq:optimal_D}.
In particular, under a variant of Assumption 1 in \cite{kaji2022metropolis} (rephrased as Assumption \ref{ass:kaji1}  in the Supplement)  the following statement holds for the classification-based estimator $\hat D_T$  defined in \eqref{eq:estimator}. For a non-negative sequence $\delta_{T,n}$  determining the quality of the training algorithm for the discriminator we have
\begin{equation}
d_H(\hat D_{T}, D^*)=O_{\bar P}(\delta_{T,n}),\label{eq:delta}
\end{equation}
where $\bar P$ encompasses all randomness in the reference table ($T$ pairs of $n$-dimensional vectors of simulated data from $M_1$ and $M_2$)   for obtaining $\hat D_T$. Recall that $d_H(\cdot,\cdot)$ was defined in the notation section as a ``Hellinger-type" metric.
Here, we establish consistency of the Bayes factor estimator under the same assumptions, additionally requiring our Assumption \ref{ass:KL}.  It will  be convenient to work with the natural logarithm of the Bayes factor, referred to by I. J. Good as the weight of evidence. 

\begin{lemma}\label{thm:consistency_BF}
Assume that a classifier $\hat D_T$ in \eqref{eq:estimator} has been trained on $T$ pairs of training data from $M_1$ and $M_2$.   Under the Assumptions 
\ref{ass:kaji1} and \ref{ass:kaji2} in the Supplement  we have for $j=1,2$
$$
E_j\left| \log \widehat{BF}_{1,2}(Y)-\log  {BF}_{1,2}(Y)\right|=O_{\bar P}( \delta_{T,n}).
$$
 \end{lemma}
\proof Section \ref{sec:proof_lemma}

 Note that the expectation in Lemma \ref{thm:consistency_BF} is with respect to the {\em marginal} distribution $\pi(\cdot\C M_1)$ as opposed to the product measure $P^{\infty}_*$. 
Lemma \ref{thm:consistency_BF} implies certain ``consistency" of our estimator in the event that $ \delta_{T,n}\rightarrow 0$ for any $n$ as $T\rightarrow\infty$, i.e. when the class of discriminators $\mathcal D_n$ is rich enough to be able to approximate $D^*$ sufficiently close. 
Our approximability/closeness assumption is stated   implicitly inside Assumption \ref{ass:kaji1} in the Supplement. For a class of classifiers $\mathcal D_n$, depending on the nature of $D^*$ (e.g. smoothness), we find the smallest $\delta_{T,n}$ for which this assumption is satisfied.  Unlike with other approximability results for deep learning (see \cite{johannes} and references therein),  we do not have a precise characterization of $\delta_{T,n}$ in terms of the network width, depth and/or sparsity.  We do acknowledge that  there is still a gap between  theory and practice of deep learning and that our method relies on a careful choice of the deep learning architecture which may not be entirely informed by our theory.  We discuss practical aspects of picking the class of neural discriminators in Section \ref{sec:comp} (Supplement).
We  use Lemma \ref{thm:consistency_BF} as an intermediate step towards showing {\em model selection consistency} with $\log \widehat{BF}_{1,2}(Y_0^{(n)})$ shown in the following theorem. 
\begin{theorem}\label{cor:consistency}
Assume that Bayes factors are consistent at a rate $1/ \nu_n $ according to Definition \ref{def:consist} and that assumptions of Lemma \ref{thm:consistency_BF} hold. Under the Assumption \ref{ass:KL} with 
$( n^{d_1/2}\vee{n^{d_2/2})  \delta_{n,T}\bar \delta_n}=o(1/\nu_n)$ for some arbitrarily slowly increasing sequence $\bar\delta_n>0$ we have
$$
\lim_{n\rightarrow\infty} P^\infty_* (\log \widehat{BF}_{1,2}(Y_0^{(n)}) <1/(2\nu_n))=0\quad\text{when $p^*\in\mathcal F_0$} 
$$
and
$$
\lim_{n\rightarrow\infty} P^\infty_* (\log \widehat{BF}_{1,2}(Y_0^{(n)}) >-1/(2\nu_n))=0 \quad\text{when $p^*\in\mathcal F_1$}. 
$$
\end{theorem}
\proof See Section \ref{pf:consistency}.

Theorem \ref{cor:consistency}  shows that even if the Bayes factor estimator is inconsistent (i.e. $  \delta_{T,n}$ {\em does not} vanish as $T\rightarrow\infty$), Deep Bayes Factors can still be consistent (in terms of model choice) as long as the true Bayes factors are consistent and   $(n^{d_1/2}\vee n^{d_2/2})\delta_{T,n}$ is slower than the speed $1/\nu_n$ of Bayes factor divergence. Now, we review sufficient conditions under which the Bayes factor (and also our estimated Bayes factor) are, in fact, consistent for model selection when $p^*\in\mathcal F_0$. The case   $p^*\in\mathcal F_1$ is symmetrical. These conditions essentially correspond to \cite{tokdar2010bayesian} and \cite{mcvinish2009bayesian}.
\begin{ass}\label{ass:conc}
 Assume that there exists a metric $d(\cdot,\cdot)$  and $\theta^*\in R^{d_1}$ such that
$$
\pi(\theta_1: d( \theta_1,\theta^*)< \varepsilon_n \C Y_0^{(n)}, M_1)\rightarrow 1\quad\text{ in $P^{\infty}_*$-probability}
$$
for some $ \varepsilon_n>0$.  At the same time, assume that  such $\varepsilon_n$ satisfies for some $\eta_n\rightarrow 0$
$$
\pi_1 (\theta_1: d( \theta_1,\theta^*)<  \varepsilon_n)=O(n^{-d_2/2}\e^{-1/\eta_n}).
$$
\end{ass}
This assumption sets a limit to the amount of posterior updating of a prior support of a neighborhood of $\theta^*$ for model $M_1$. 
Note that when $p^*$ does not belong to the family $\mathcal F_1$, $\theta^*$ can be the  point corresponding to the KL projection of $p^*$ onto $\mathcal F_1$.
 \begin{theorem}\label{thm:convP}
 Under the Assumptions of Lemma \ref{thm:consistency_BF}, Assumption \ref{ass:KL}  and  Assumption \ref{ass:conc}, as long as  $n^{d_2/2}\bar \delta_n \delta_{n,T}=o(1/\eta_n)$ for some arbitrarily slowly increasing sequence $\bar\delta_n>0$,  we have for $p^*\in \mathcal F_1$
 $$
 \lim_{n\rightarrow\infty}\widehat{BF}_{1,2}(Y_0^{(n)})=0\quad\text{in $P^\infty_*$-probability}.
 $$
 \end{theorem}
\proof See Section \ref{pf:convP}.
\begin{remark}
The actual Bayes factor ${BF}_{1,2}(Y_0^{(n)})$ converges to  $0$ in $P^\infty_*$-probability under Assumption  \ref{ass:KL}(b) and  Assumption \ref{ass:conc} (according to Theorem 4.1 in \cite{tokdar2010bayesian}. 
A similar conclusion holds for $\widehat{BF}_{2,1}(Y_0^{(n)})$ within the context of Theorem \ref{thm:convP} with the assumptions about priors $\pi_1$ and $\pi_2$ and models $M_1$ and $M_2$ reversed.
\end{remark}

\section{Generative AI for Bayesian Model Criticism}\label{sec:gAI}
 
Because a model can be criticized without an explicit consideration of an alternative  \citep{gelfand1992model}, it is difficult to entirely separate model selection from
assessments of model adequacy.   While our paper focuses on {model} selection, in this section we partially address  how  generative  contrastive approaches could be used for model criticism.
One of the key tools for Bayesian model criticism is the posterior predictive check (PPC) of \cite{guttman1967use} whose premise is: ``If my model is good, then its posterior predictive distribution will generate data that looks like my observations". In PPC, the closeness is typically assessed through a diagnostic function, and a reference distribution has to be chosen to determine whether this diagnostic function indicates extreme values.  An example diagnostic is the average squared distance to the posterior mean $ \frac{1}{n}\sum_{i=1}^n(Y_i^0 -   E[Y_i\C Y_0^{(n)}])^2$. A common criticism of such approaches is the double use of the data $Y_0$, for the reference distribution and for the evaluation of the diagnostic function. Avoiding the double of use of data, \cite{bayarri2000p} propose the partial predictive check. \cite{johnson2007bayesian} proposed to use pivotal diagnostic to ensure the diagnostic and reference distributions are uncorrelated. \cite{moran2019population} considered population predictive checks based on data splitting. 
Below, we outline certain approaches which may serve as an alternative to these developments.

 Assume that $Y_0^{(n)}$ is a vector of $n$ independent observations. Suppose that now we train a classifier $\hat d$ contrasting the {\em observed data} $Y_0^{(n)}$ with fake simulated data {$\tilde Y = (\tilde Y_1,...,\tilde Y_n)'$, where $\tilde Y_i\overset{i.i.d.}{\sim} \pi(\cdot\C Y_0, M)$ are  generated from the posterior predictive distribution under the hypothesized model $M$.} Note that this is very different from the setup in Section \ref{eq:our_method} where we were contrasting two fake datasets from models $M_1$ and $M_2$. If the observed data are indistinguishable from the simulated  data, the estimated classifier should satisfy 
\begin{equation}\label{eq:Z_dist}
Z(\tilde Y,\hat d)\equiv\frac{1}{n} \sum_{i=1}^n   {\hat d(\tilde  Y_i)}\approx \frac{1}{2}.
\end{equation}
During the training of  $\hat d$, the observed data $Y_0^{(n)}$ is also used twice (first to simulate the fake data and then to train the classifier). 
However, this seems different from the usual double use of the data where $Y_0^{(n)}$ is used to  construct a reference distribution and to evaluate the test statistic inside this distribution. 
Here, we can evaluate the test statistic  $Z$ on the fake data $\tilde Y$, not the observed data $Y_0^{(n)}$.  Instead of a ``p-value" style measure of surprise, one can sample from the distribution
$\pi(Z(\tilde Y,\hat d)\C Y_0^{(n)})$ by evaluating $Z(\tilde Y,\hat d)$ on new plugged-in samples $\tilde Y\sim\pi(Y\C Y_0^{(n)}, M)$  from the posterior predictive distribution. 
Whether or not the model $M$ is good could be determined from the concentration of this distribution around $1/2$.
Sampling from the parameter posterior using generative techniques \citep{wang2022adversarial,polson2023generative} makes sampling from $\pi(Z(\tilde Y,\hat d)\C Y_0^{(n)}, M)$ feasible. To alleviate  systematic bias of a pre-trained classifier $\hat d$, one might need   retrain    $\hat d$ for each replication of $\tilde Y$. This strategy might be a new promising route towards implementing model criticism, i.e. compatibility of observed data with predictions from a model.  {An example of this idea is presented in Section \ref{toy_continue_3} in the supplement.}

We conclude this section with further discussion of possible model criticism developments using generative Bayes in the context of existing literature.
 \cite{gelfand1992model} and \cite{box1980sampling} argue that model criticism  should be addressed using posterior predictive distributions which are consonant with the intended use of the model.
In addition, predictive distributions are compatible across models while posteriors are not.   The methodology of \cite{gelfand1992model} rests upon the ability to obtain desired predictive distributions and to calculate expectations under these distributions.  These authors argue for a cross-validation viewpoint, comparing an observed value against  conditional predictive distributions arising from single point deletions. In particular, one of their suggestions is to construct a discrepancy measure (checking function) comparing each observation $Y_i^0$ against a posterior expectation $E[Y_i^0\C Y_{\backslash i}^0]$ where $Y_{\backslash i}^0$ are all but the $i^{th}$ observation inside $Y^{(n)}_0$. The conditional expectation after removing the observation $Y_i^0$ is not too easy to compute and they suggested importance sampling. We suggest one feasible computational strategy based on generative posterior samplers. Using the Bayesian GAN approach \citep{wang2022adversarial}, one can train a GAN mapping $g(Y_{n-1},X)$ to simulate from $\pi(\theta\C Y_{n-1})$ by passing a random noise variable $X$ through this mapping for a fixed vector $Y_{n-1}$  of length $ n-1 $. This training is performed solely on fake data.  One can then plug a vector $Y_{\backslash i}^0$ into $g(Y_{\backslash i}^0,X)$ to perform iid sampling from the posterior distribution $\pi(\theta\C Y_{\backslash i}^0)$ for any $i=1,\dots, n$. This sampling is cost-free because it is merely based on passing noise through the trained mapping evaluated at $Y_{\backslash i}^0$. 
The ability to simulate from this posterior  makes simulating from $\pi(Y_i\C Y_{\backslash i}^0)$ and, therefore, approximating $E[Y_i\C Y_{\backslash i}^0]$ feasible.
One can, for example,  compute the discrepancy measure $Z(Y_0)=\frac{1}{n}\sum_{i=1}^n (Y_i^0-E[Y_i\C Y_{\backslash i}^0])^2,$ which avoids the use of observation $Y_i^0$ for the predictive check on $Y_i^0$. A tail area evidence measure can be then obtained by assessing the location of $Z(Y_0)$ relative to the a reference distribution of $Z(Y)$ where, for example, $Y$ has been simulated from the (posterior predictive under) model $M$.

\section{Evaluation Toolkit}\label{sec:numerical}

 We compare the  performance of Deep Bayes Factors with several existing Bayes factor estimators using various  estimation and model inference metrics. 

\subsection{Estimation Consistency} 
To capture the inclination of Deep Bayes Factors to achieve \emph{estimation consistency}, we compute the MSE of the log Bayes factor defined by 
\begin{equation*}\label{eq:mse_msr}
    MSE_{logBF} = \sum_{j=1}^2 \pi(M_j)~\hat{\E}_{Y\sim M_j}\left[\log {BF}_{1,2}(Y)-\log\widehat{BF}_{1,2}(Y)\right]^2,
\end{equation*}
where $\hat{\E}_{Y\sim M_j}$ is the empirical expectation over $T_0=1,500$ draws from a model $M_j$. {Note that this criterion clearly mirrors the Bayes factor consistency in Lemma \ref{thm:consistency_BF} and is therefore expected to be small  for a flexible enough  neural network class}. A simple scatterplot of estimated $\widehat{BF}_{1,2}(Y)$ versus true Bayes factors $BF_{1,2}(Y)$ on data samples $Y$ obtained from $M_1$ and $M_2$ can reveal proclivities to under/over-state model evidence. Figure \ref{fig:scatterplot} shows cartoons of plausible scatterplot behaviors.   Points in the first and third quadrant centered at (1,1)  indicate that the binary decision based on the decision threshold $1$ leads to the same conclusion.

\begin{figure}[!t]
    \centering
    \includegraphics[width=\linewidth]{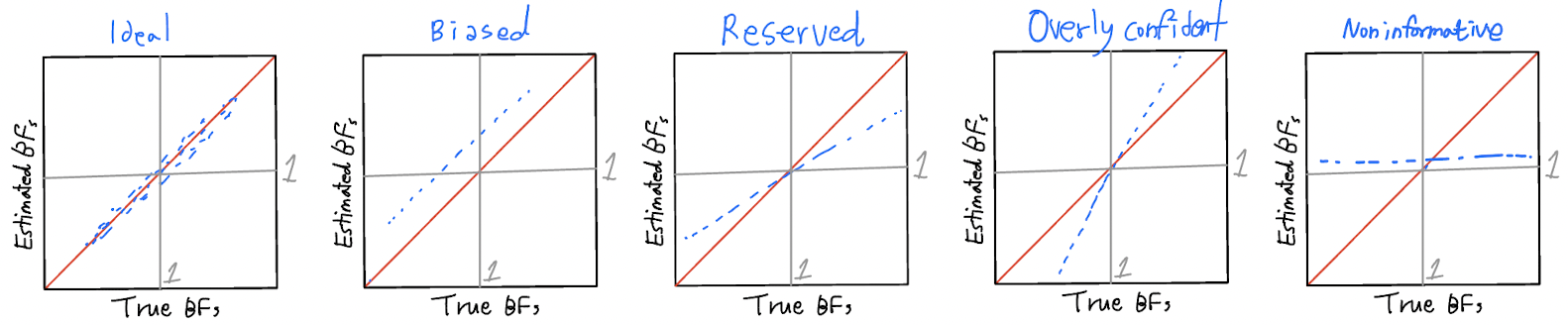}
    \vspace{-1cm}
    \caption{Interpretation of   Bayes factor scatterplots}
    \label{fig:scatterplot}
\end{figure}

The strength of our generative approach to Bayes factor estimation is that we can evaluate the estimator at any data $Y$, not just the observed data $Y^{(n)}_0$. This allows us to recreate distributions of Bayes factor estimators when $Y$ is generated from either model $M_1$ and model $M_2$.  Inspecting entire {\em distributions} of Bayes factors $BF_{1,2}(Y)$   provides a more comprehensive image of discriminability among the two models. The quality of our estimator can be gauged by the closeness of the distributions of true and estimated Bayes factors. Figure \ref{fig:four_types} {in the supplement} portrays cartoons of plausible shapes of   Bayes factor distributions (true and estimated) under the two models. The relative position of the true (or estimated) $BF_{1,2}(Y_0)$ within the plot provides  stronger information than merely deciding whether the Bayes factor is above or below $1$. In addition, we  estimate the KL divergence between empirical distributions (kernel density estimates (KDE)). These distances  are computed separately under each model and then subsequently weighted by the prior model probability. {Similarly, we measure the Spearman's rho correlation (rank-based) between the estimated and true BF's for simulated data under each model and take their weighted (by the model prior probability) sum}.
 
 We also compute an unbiased estimator of the prior $\pi(M_j)$ by averaging $\pi(M_j\C  Y)$ over the simulated data $(Y_1,...,Y_{T_0})'\sim P(Y)$, denoted as $\hat{\pi}(M_j) = \frac{1}{T_0}\sum_{i=1}^{T_0} \pi(M_j\C  Y_i).$ If $\hat{\pi}(M_j)$ deviates significantly from true $\pi(M_j)$, the estimator may not be trusted \cite{schad2022workflow}. In our numerical examples, we assume $\pi(M_1)=\pi(M_2) = 0.5$.

\subsection{Measures of Surprise}
Our simulation-based framework lends itself naturally to {\em inference} with Bayes factors based on tail-area surprise measures.
While the $p$-value is a tail area (as opposed to likelihood-based) evidence measure, it can be calibrated to obtain a lower bound to a Bayes factor \citep{bayarri1998robust}. Before seeing the data, Bayes factors are random variables and we can think of them as a proxy for a test statistic.  Evaluating the position of the Bayes factor $BF_{1,2}(Y_0^{(n)})$ within the distributions of $BF_{1,2}(Y)$ for $Y\sim \pi(\cdot\C M_1)$ and  $Y\sim \pi(\cdot\C M_2)$ has a flavor of 
the prior predictive $p$-value popularized by \cite{box1980sampling} .  
We consider surprise measures
   \begin{equation}
  p_1(Y_0^{(n)})= P(B_{1,2}(Y)> B_{1,2}(Y_0^{(n)}) \C M_1)\quad\text{and}\quad p_2(Y_0^{(n)})=P(B_{1,2}(Y)\leq B_{1,2}(Y_0^{(n)})\C M_2),\label{eq:tails}
   \end{equation}
    which are essentially prior predictive $p$-values using the Bayes factor as a departure statistic. We do not argue that the prior-predictive reference distribution is   preferred over alternatives including the conditional predictive $p$-value \citep{bayarri1998robust} which is justifiable when there is suspicion that poor priors were used and which is provably superior to various Bayesian $p$-values in large samples \citep{robins2000asymptotic}. In simulation-based models  considered here, however, it might be difficult to come up with appropriate statistics needed for the computation of the conditional predictive $p$-values. While our computational framework enables the use of posterior-predictive $p$-value \citep{guttman1967use,rubin1984bayesianly} by simulating $Y$ from the posterior (not prior) predictive under each mode, we focus on the prior predictive $p$-values acknowledging its weaknesses in its prior sensitivity.
{Using surprise measures for Bayes factors as inferential tools} is not a new concept. For example, \cite{vlachos2003calibration} suggest  a simulation-intensive approach to calculate various tail area measures of surprises for Bayes factors  and \cite{garcia2005calibrating} propose  a Monte Carlo method to find a decision rule $c$ such that $P(B_{1,2}(Y)\leq c|M_1)=P(B_{1,2}(Y)\geq c|M_2)$. Computing such surprise values was also considered in \cite{schonbrodt2018bayes} who proposes a Bayes factor design analysis aimed at determining an appropriate sample size $n$ that ensures that with a high probability the sample data from the models exhibit strong Bayes factors in favor of the true model.
 When closed-form evaluations of Bayes factors are unavailable, the computation of surprise measures   can be  involved. The ability to rapidly evaluate the Deep Bayes factor mapping for any $Y\in{\rm Supp}(P)$ facilitates the computation of surprise measures.  For example, we approximate $ p_1(\cdot)$ and $ p_2(\cdot)$ in \eqref{eq:tails} by replacing $BF_{1,2}(\cdot)$ with an estimator $\widehat{BF}_{1,2}(\cdot)$  and by replacing tail probabilities with empirical proportions based on simulated data $Y$. {We denote these approximations by $\hat p_1(\cdot)$ and $\hat p_2(\cdot)$.} Since the inferential quality of a Bayes factor estimator can be assessed by how well it approximates the tail probabilities in \eqref{eq:tails}, we measure the MSE between the actual surprise and the estimated surprise by 
   $ MSE_{surp} = \sum_{j=1}^2 \pi(M_j)~\hat{\E}_{Y\sim M_j}\left[ p_j(Y)-\hat p_j(Y)\right]^2.$

\subsection{Model Inference}\label{sec:minf}
 The actual precision in estimating   Bayes factors may be less relevant  for decision making when there is overwhelming evidence for one of the two models (and the estimation error is appropriately small). It is the quality of decisions with Bayes factor's that  ultimately matters, not the approximation quality of its estimator. Jeffreys' Bayes factor scale \citep{jeffreys1998theory} is a suggested interpretation of the evidence strength based on the magnitude of the Bayes factor binned into  10 categories. The coarsest categorization is into two classes, based on the decision boundary at  $1$ (values greater than one favor model $1$ and values smaller than one favor model $2$). 
 To quantify the inference quality for various thresholds (not just $1$), we use the area under the ROC curve (AUC) computed as follows. Considering $1$ as a ``positive" label and $2$ as  ``negative", we can calculate the true positive rate (TPR) and the false positive rate (FPR). Given the decision threshold $c\in(0,\infty),$ we write TPR and FPR as $TPR_c(\widehat{BF}_{1,2})=\mathbb{P}(\widehat{BF}_{1,2}(Y)>c|Y\sim M_1)$ and $FPR_c(\widehat{BF}_{1,2})=\mathbb{P}(\widehat{BF}_{1,2}(Y)>c|Y\sim M_2)$, where $\mathbb{P}$ is the empirical distribution of a simulated sample from the model. As these values depend on the choice of $c$, we can draw the ROC curve representing the relationship between TPR and FPR when  $c$ is sent from 0 to $\infty$. The AUC  does not penalize biased or improperly scaled estimation.}

\section{Simulation Study}\label{sec:sim}

We apply Deep Bayes Factors to several examples for which true Bayes factors are known but are also known to be difficult to estimate.

\subsection{Toy Examples}\label{sec:simul} 
{In this section, we focus on the Negative Binomial vs. Poisson distributions in Example \ref{ex:toy}. {For the other two datasets, see Section \ref{sec:other_datasets}}. This dataset can represent various scenarios with different hyperparameters $(\alpha_1,\beta_1,\alpha_2,\beta_2)$, which determine the landscape of the decision boundary of the model selection problem. Here, we choose $(\alpha_1,\beta_1,\alpha_2,\beta_2)= (2,2,4,4)$, which we call Dataset 1 (i), and $(\alpha_1,\beta_1,\alpha_2,\beta_2)= (1,1,1,1)$ as in Example \ref{ex:toy}, which we call Dataset 1 (ii). The former is more separable near decision boundaries than the latter in the sense that the probability that $BF_{1,2}(Y)$ is near $1$ is smaller (see, Figure \ref{fig:separability} in Section \ref{sec:additional_plots}). The true Bayes factor is available in a closed form \citep{BERN}.} 
Letting \(Y_\cdot = \sum_{i=1}^nY_i\),
\begin{equation*}
    BF_{1,2}(Y_1,Y_2) = \frac{\Gamma(\alpha_1+\beta_1)\Gamma(n+\alpha_1)\Gamma(Y_\cdot+\beta_1)\Gamma(\alpha_2)\Gamma(n+\beta_2)^{Y_\cdot+\alpha_2}\prod_{i=1}^2Y_i!}{\Gamma(\alpha_1)\Gamma(\beta_1)\Gamma(2+Y_\cdot+\alpha_1+\beta_1)\Gamma(Y_\cdot+\alpha_2)\beta_2^{\alpha_2}}.
\end{equation*} 
Because of the explicit likelihood, we can readily use the well-known Bridge sampling \citep{bennett1976efficient,meng1996simulating} and Warp Bridge sampling \citep{meng2002warp} as baselines. {We set the number of MCMC iterations of Bridge and Warp sampling as 150,000 with the burn-in  10,000.} We also evaluate the ABC-based Bayes factor estimator \citep{robert2011lack}, which is commonly employed in applications without an explicit likelihood. To compute the ABC-based estimates for a large number of simulated samples within a reasonable timeframe, we adopt a rank-based rule for the ABC algorithm instead of the standard \( \epsilon \)-acceptance rule. This approach allows for the reuse of ABC samples across different simulated datasets simultaneously, significantly reducing computation time. {The detailed algorithm is provided in Section \ref{sec:ABC_}, and hyperparameter choices are listed in Table \ref{tab:ABC_spec} therein. Consequently, the top 120 ABC samples are used to compute the ABC-based estimator in \eqref{eq:abc1} with a small adjustment to avoid numerical explosion. As a result, the range of the ABC-based estimators is approximately [1/120,120].} We vary the sample size from \(n=2^1\) to \(n=2^7\), specifically focusing on sample sizes under 1,000. Studying such small sample sizes remains relevant for practical purposes, e.g., in cognitive and psychological science, sample sizes often range from 30 to 60 \citep{schad2022workflow} and a sample size of 180 is considered large \citep{schad2022workflow}. For each \(Y_0^{(n)}\), we run our algorithm and the baseline methods 20 times to assess estimation stability. {As discussed in Section \ref{sec:comp}, we use both FNNs and BNNs for our algorithm, which we will refer to as DeepBF-FNN and DeepBF-BNN, respectively, {where BNNs stand for batch normalization neural networks}. 

{We begin by assessing estimation consistency, as illustrated in Figures \ref{fig:kde} and \ref{fig:consis2} (Section \ref{sec:additional_plots}) for a sample size \( n = 128 \). The first row of these figures presents scatter plots that compare the estimated \(\widehat{BF}_{1,2}(Y)\) with the true Bayes factors \(BF_{1,2}(Y)\) from simulated samples \( Y \). The scatter plot for DeepBF-BNN suggests unbiased and properly scaled estimates, while DeepBF-FNN exhibits more reserved scaling. In Figure \ref{fig:consis2}, both Bridge and Warp methods yield well-scaled and unbiased estimates. In contrast, Figure \ref{fig:kde} shows  bias in both methods, although biased estimates may still hold utility within a rank-based decision rule, as reflected by the surprise measures in \eqref{eq:tails}. The ABC method appears non-informative. Additional insights are offered in Figures \ref{fig:scat2} and \ref{fig:scat3} (Section \ref{sec:additional_plots}), which present scatter plots for increasing values of \( n \). Here, the ABC method provides more accurate estimations for smaller \( n \) values, but these estimations become non-informative as \( n \) increases. An increase in the number of ABC samples, currently at 1,200,000, could potentially improve these outcomes. On the contrary, Bridge and Warp show unstable estimates for \( Y \sim M_2 \) but improve as \( n \) becomes larger. While DeepBF-FNN tends to outperform DeepBF-BNN, its efficacy wanes with larger \( n \), possibly due to the network's fixed intermediate width across all \( n \) values to stabilize optimization (Section \ref{sec:comp}).

{\spacingset{1.5}
\begin{figure}
    \centering
    \includegraphics[width=0.95\linewidth]{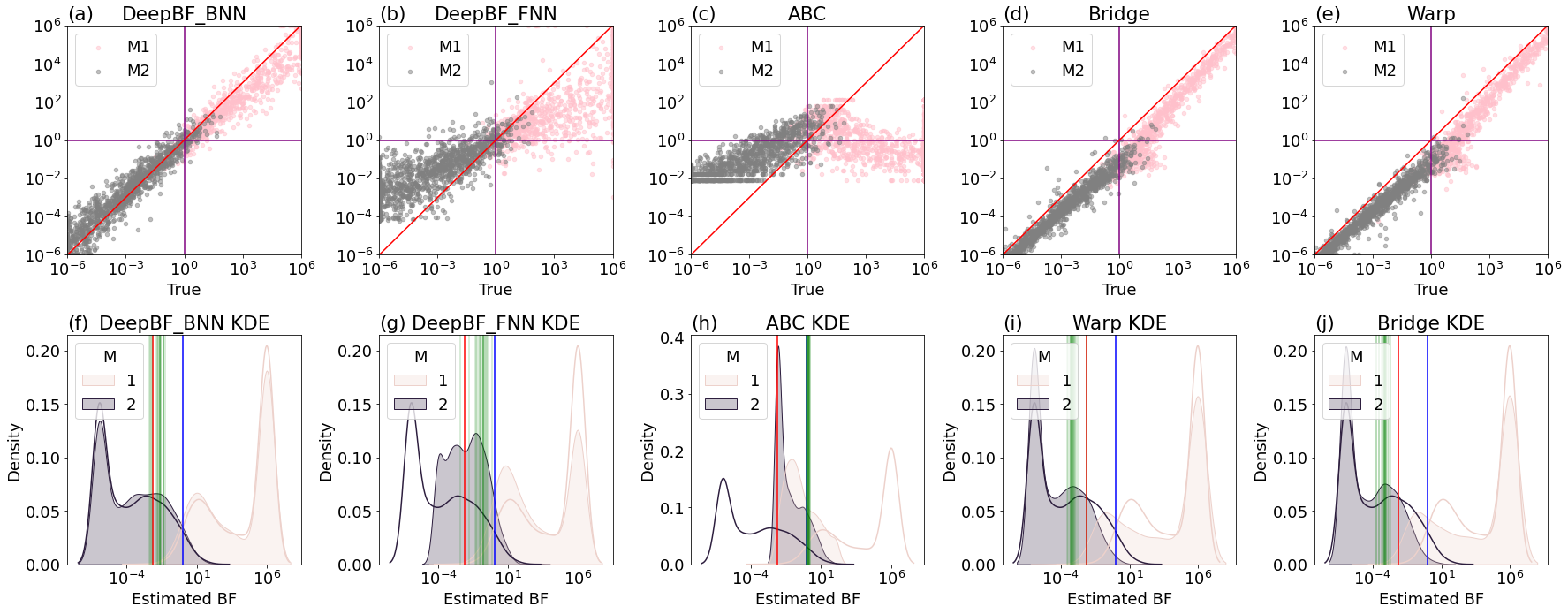}
  \caption{Data 1 (i):  Negative Binomial vs. Poisson distributions when $n=128$. First row: Scatter plots of estimated vs. true BF's. Second row: KDE plots on estimated BFs (colored regions) and on true BFs (lines). The red line marks \( BF_{1,2}(Y_0^{(n)}) \) for one particular $Y_0^{(n)}$ from $M_2$, green lines denote \( \widehat{BF}_{1,2}(Y_0^{(n)}) \) over 20 trials, and the blue line is at the decision threshold of 1. x-axis:  values thresholded to the range \( [10^{-6},10^6] \). } 
    \label{fig:kde}
\end{figure}
}
In the second row of Figure \ref{fig:kde} and \ref{fig:consis2} (Section \ref{sec:additional_plots}), we have empirical distribution plots of the Bayes factors (true and estimated) under the two models. While the KDE shapes of these plots corroborate the scatter plot shapes, we can also see the separability of distributions under each model, and the overall distributional deviation of the estimates. We also visualize the estimation quality for an instance $Y_0^{(n)}$ drawn from $M_2$. In Figure \ref{fig:kde} (f-j), each vertical red line represents the true $BF_{1,2}(Y_0^{(n)})$, and the green lines represent its estimates from 20 different trials. For DeepBF-BNN, the green lines cluster closely around the red line, whereas for all other methods, the green lines show more deviation. The blue line is a reference line at the decision threshold of 1. Our algorithm's estimates tend to be more conservative, closer to the blue line, while the Warp and Bridge estimates are overly optimistic, diverging further from the blue line. The variance in the multiple estimates increases in the following order ABC $<$ Warp and Bridge $<$ DeepBF-BNN $<$ DeepBF-FNN. More comprehensively, the scatter plots among multiple DeepBF{-BNN} estimates are presented in Figure \ref{fig:bnn-consist} (Section \ref{sec:additional_plots}). The plots in Figure \ref{fig:consis2} (f-j) display similar results, but with Warp and Bridge estimates appearing unbiased and clustering tightly around the true red line. 

{\spacingset{1.5}
\begin{figure}
    \centering
    \includegraphics[width=\linewidth]{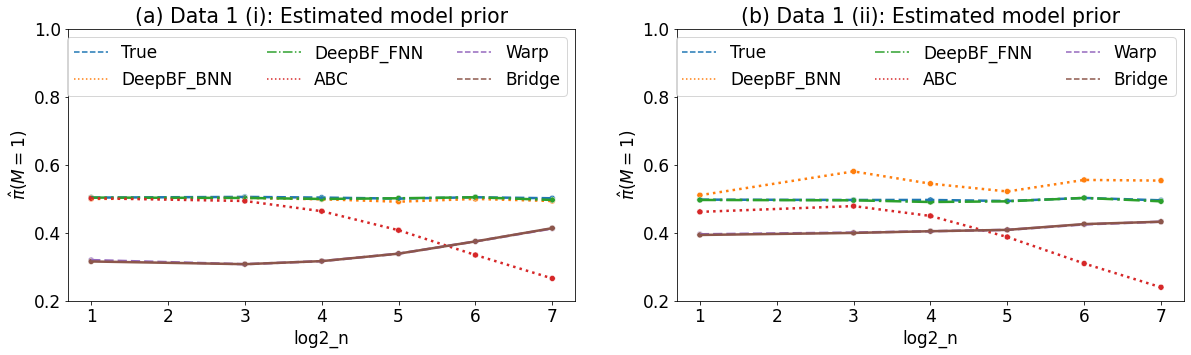}
    \vspace{-1cm}
\caption{Estimated model priors over increasing \( n \); ideal value is 0.5. (a) Data 1 (i), (b) Data 1 (ii)}
    \label{fig:half}
\end{figure}
}

In Figure \ref{fig:half} (a), the estimated model priors are presented. While the Warp and Bridge methods improve as $n$ increases (as the problem becomes more binarized), the ABC method exhibits a gradual decline in performance, and DeepBF-FNN is relatively more stable than DeepBF-BNN. For Data 1 (ii), similar results are observed as in Figure \ref{fig:half} (b). Figure \ref{fig:three_col} shows the Spearman's rho correlation (rank-based), the empirical KL divergence between the true and estimated distributions, and the MSE of log BFs. Consistent with earlier observations, we see varying trends as $n$ increases: Spearman's rho indicates an improvement in rank-based correlation for the Warp and Bridge methods, a decline for the ABC method, and initially an outperformance by DeepBF-FNN, which is eventually surpassed by DeepBF-BNN at higher $n$. The empirical KL divergence and MSE metrics generally corroborate these trends, reinforcing the comparative assessment of these methods. Lastly, to evaluate inferential consistency, ROC curves and corresponding AUC values derived from the true and estimated Bayes factors are presented in Figure \ref{fig:roc} (a-b) {for Data 1 (i)}. The estimates based on DeepBF surpass those from Warp and Bridge at \( n=2 \), and are comparable to these methods at \( n=128 \). Figure \ref{fig:roc} (c) illustrates the MSE for surprise measures, with DeepBF-FNN and DeepBF-BNN showing superior consistency across varying \( n \) values compared to other benchmarks. Similar results are shown in Figure \ref{fig:roc} (d-f) on Dataset 1 (ii). {These results demonstrate that DeepBF reliably approximates distributional inference metrics like surprise, with the added convenience of simultaneous computation.} In conclusion, DeepBF demonstrates its competitiveness, even without the provision of explicit likelihood, unlike Warp and Bridge sampling.

{\spacingset{1.5}
\begin{figure}[!t]
    \centering
    \includegraphics[width=0.8\linewidth]{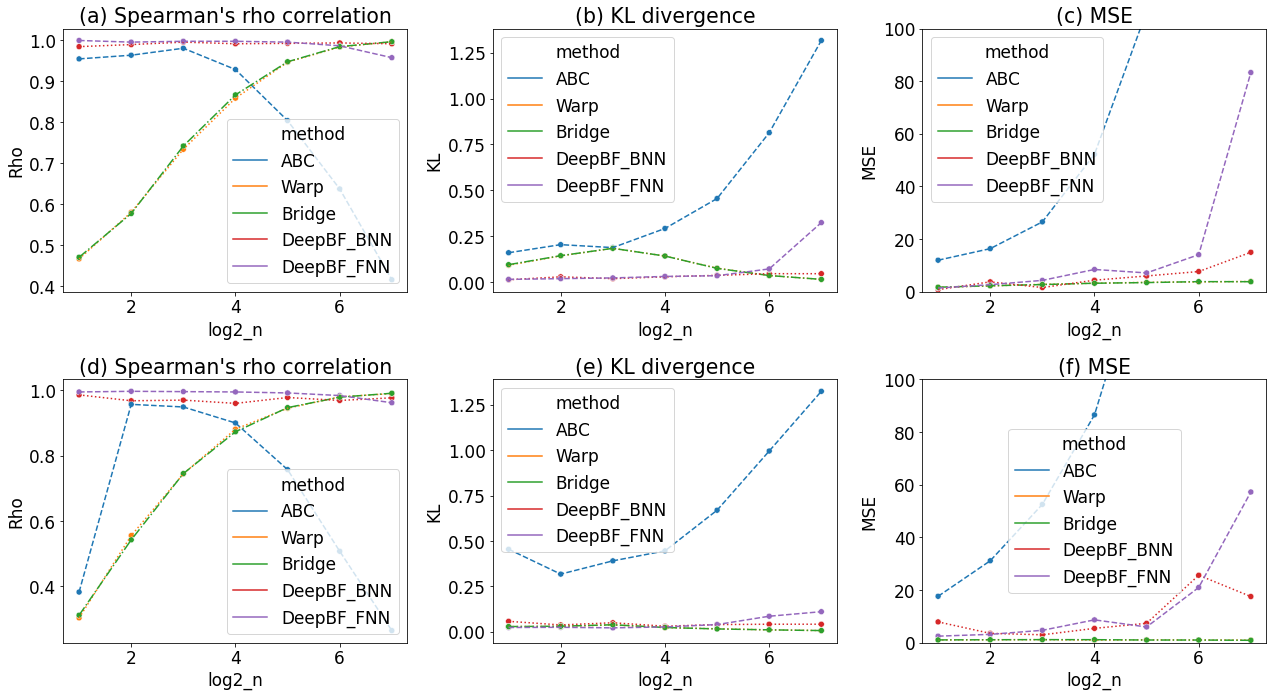}
    \vspace{-.5cm}
 \caption{Evaluation of BF approximations for Data 1 (i) in (a-c) and Data 1 (ii) in (d-f), showing Spearman's rho (a, d), KL divergence (b, e), and MSE of log BF (c, f). Higher rho and lower KL/MSE indicate better performance.}
    \label{fig:three_col}
\end{figure}
}
}

\section{Real Data Analysis -  Cognitive  Bias}\label{sec:real2}
In psychological cognitive research, the use of Bayes factors  to test among explanatory narratives has been on the rise \citep{lee2014Bayesian, heck2023review}. 
We study weapon identification task \citep{rivers2017weapons}, which delves into the impact of stereotypes on participants' ability to recognize objects. Participants were shown images and asked to distinguish between tools and weapons. Each image presented either a tool or a weapon alongside one of three priming images: a white face, a black face, or a neutral face outline. The study assessed the accuracy of responses across all six combinations of primes (white, black, neutral) and targets (tool, weapon), with 36 trials for each combination, amounting to 216 trials per participant. We focus on the data subset where participants were given a 1,000 ms response deadline. This subset corresponds to a group of  $N=42$ participants. 

{\spacingset{1.5}
\begin{figure}
    \centering
    \includegraphics[width=.9\linewidth]{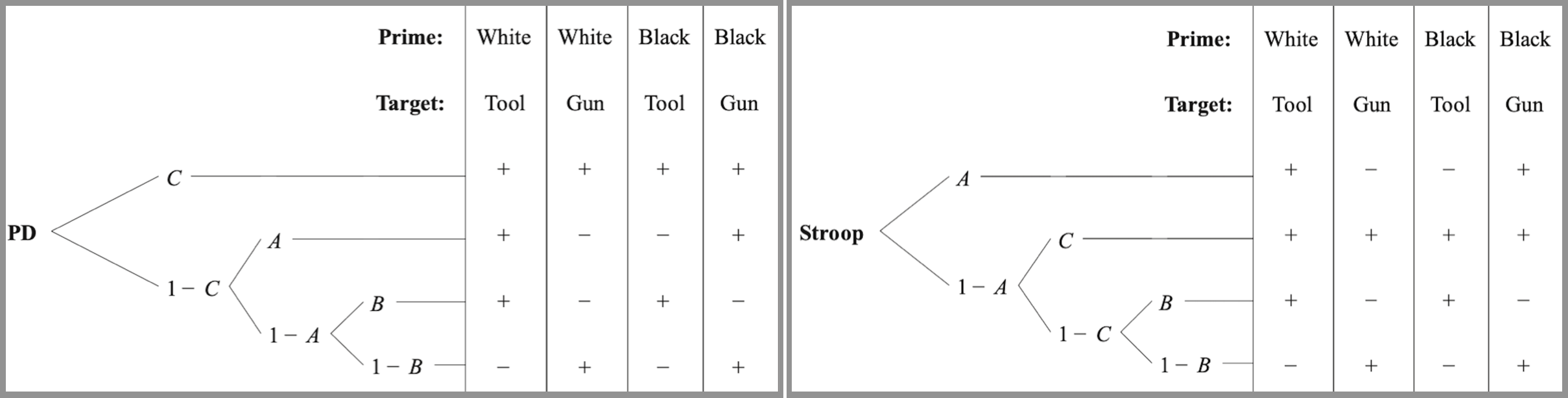}
    \vspace{-.5cm}
    \caption{Figure captured from \cite{heck2023review}. Each branch of the probability tree either leads to a correct (+) or incorrect (-) response. The parameters refer to the probabilities that one of the hypothesized processes succeeds (C = controlled; A = automatic; B = guessing Tool).}
    \label{fig:weapon}
\end{figure}
}
For this task, \cite{bishara2009multinomial, heck2023review} compared two multinomial-processing tree models: $M_1$, the process-dissociation model with guessing (PD), and $M_2$, the Stroop model with guessing (Stroop). {Both models are depicted in Figure \ref{fig:weapon}, each with a different order of cognitive processes. In essence, testing between the two models amounts to determining whether stereotypes (prejudice) are activated before a random guess in the recognition of weapons.} In the diagrams, \(A\) denotes the probability of automatic stereotype activation, and \(B\) stands for the likelihood of a random guess. \(C\) represents the probability that the controlled process is active. In the PD model, controlled processing probability \(C\) is considered first. If controlled processing is not active (with a probability of \(1-C\)), an individual might rely on their automatic stereotype with probability \(A\). If not, they might make a random guess with probability \(B\). Conversely, in the Stroop model, the automatic stereotype is activated before the controlled process. In \cite{bishara2009multinomial, heck2023review}, the total number of successes for each participant follows a binomial distribution, denoted as $Bin(36,p_{j})$, where $j=1,...,6$ represents the six prime-tool combinations. The corresponding values for \(p_{j}\) under each model are presented in Table \ref{tb:pij} in Section \ref{sec:additional_plots}.

{\spacingset{1.5}
 \begin{figure}
    \centering
    \includegraphics[width=.85\linewidth]{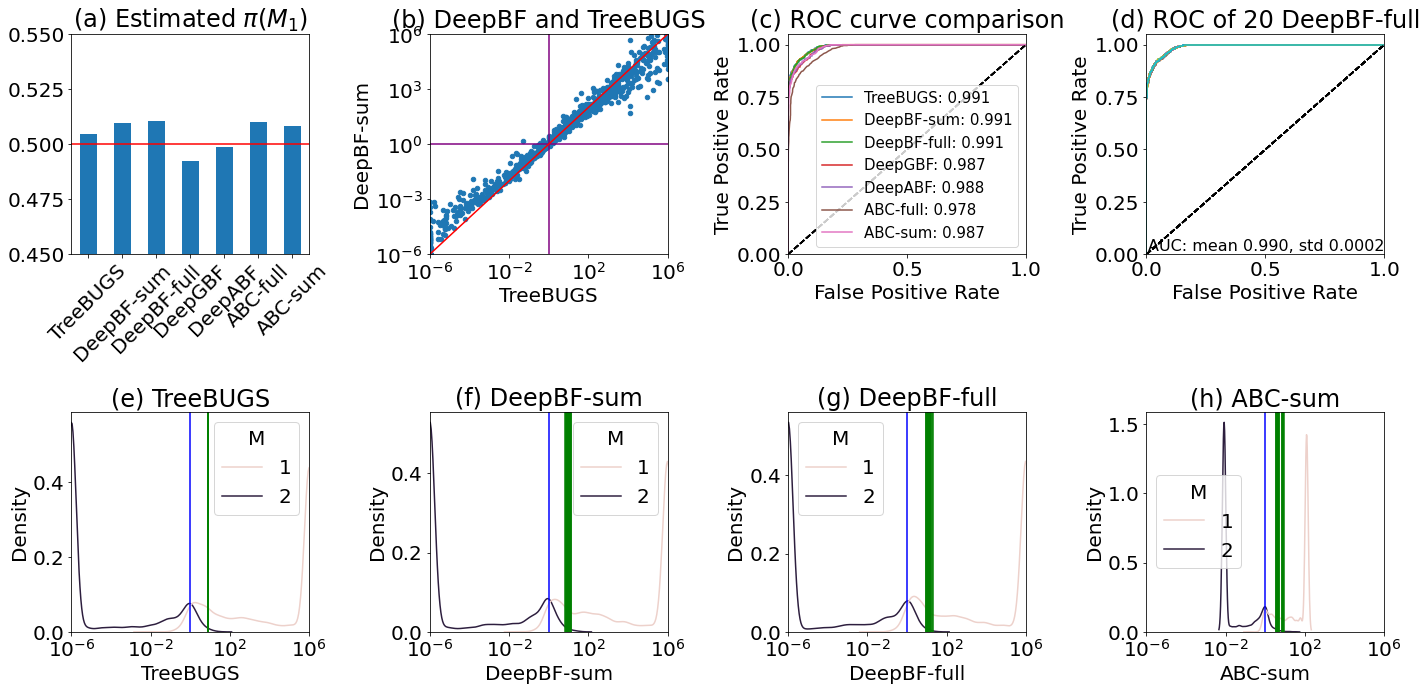}
    \vspace{-.5cm}
    \caption{(a) The estimated priors. True is $\pi(M_1)=0.5$. (b) The scatter plot of the BF estimates by DeepBF and TreeBUGS. (c) The ROC curves of methods in comparison (d) The ROC curves of 20 DeepBF networks. (e-h) The empirical density (KDE) plot of the BF estimates with reference line at 1 (blue) and at the estimated $\widehat{BF}_{1,2}(Y_0^{(n)})$ (green). }
    \label{fig:merge5}
\end{figure}
}

As a baseline we use TreeBUGS \citep{heck2018treebugs,heck2023review}, an R package specifically crafted for hierarchical multinomial-processing tree modeling. Before the recent implementation of Bayes factor estimates inside this software, estimating Bayes factors for this type of modeling was challenging. Since TreeBUGS does not distinguish between individuals, in \cite{heck2023review}, the data was processed into a single 12-dimensional vector (6 for success and 6 for failure count), i.e., condensed as $Bin(N \times 36, p_j)$ for $j=1,...,6$. This approach inherently results in data always being represented as summary statistics.\footnote{ Otherwise, the model should be $Bin(36, p_{ij})$, with parameter $p_{ij}$ for each $i$-th participant.} {  In our approach, while we apply our method to the 12-dimensional vector, we also consider the data as $N\times 6 = 252$ dimensions, leveraging DeepBF's flexibility to treat each participant as an independent entity from \(Bin(36, p_{i})\). We denote the results as DeepBF-sum and DeepBF-full, respectively.} Since Section \ref{sec:sim} finds that a BNN is better in larger $n$ and a FNN is better in small $n$, we use a BNN for the 252 dimensional case and a FNN for the 12 dimensional case. We also compute ABC estimates (Section \ref{sec:ABC_}) for both 252 and 12 dimensional vectors with the hyperparameters in Table \ref{tab:ABC_spec}, denoting the result as ABC-full and ABC-sum, {respectively}. Lastly, we can also compute GBF and ABF values by using our DeepBF techniques, as described in Section \ref{sec:variants}. For this, considering $(n_x,n_z ) = (1, 41)$, we train an additional FNN of input size $6$ and combined with DeepBF-full.
{\spacingset{1.5}
\begin{table}
\centering\caption{The estimated Bayes factors and surprise measures on the data}
\begin{center}
\scalebox{0.9}{\begin{tabular}{c || c | c  | c |c|c|c} 
 \hline
  Method&ABC-full & ABC-sum&TreeBUGs& DeepBF-sum& DeepBF-full& DeepGBF\\
 \hline\hline
$\widehat{BF}_{1,2}(Y_0^{(n)})$ & 4.304 & 4.545&8.10 & 7.639&8.082 &  1.7\\\hline
$\hat{p}_1(Y_0^{(n)})$ &  0.301 &0.219& 0.190 & 0.184 &  0.198&0.140\\\hline
$\hat{p}_2(Y_0^{(n)})$ &  0.01 & 0.006& 0.004 & 0.005 & 0.005& 0.023
\\\hline
\end{tabular}}
\end{center}
\label{tb:troop}
\end{table}
}

{Before comparing the results from these methods, we first examine the estimated prior as a validation measure.} Across all methodologies, the estimated model priors presented in Figure \ref{fig:merge5} (a) hover around 0.5. Now, to understand the relationship between the DeepBF estimates and those obtained from other methods, we present a scatter plot for comparison in Figure \ref{fig:merge5} (b) and Figure \ref{fig:weapon_scatter} in Section \ref{sec:additional_plots}. For this purpose, we simulate 3,000 random draws from the model distribution, allocating 1,500 draws to each model. Intriguingly, the scatter plot in Figure \ref{fig:merge5} (b) highlights a strong correlation between the estimates of DeepBF and TreeBUGs. In Figure \ref{fig:weapon_scatter}, all methods are highly correlated although ABC is relatively more reserved. Again, the ROC curve for each method is illustrated in Figure \ref{fig:merge5} (c). The ROC curves of a DeepBF estimator closely resemble those of TreeBUGS. Figure \ref{fig:merge5} (d) displays the ROC curves of 20 independently trained DeepBF estimators, all of which are highly similar. To the actual dataset $Y_0$, we apply each method 20 times to compute $\widehat{BF}_{1,2}(Y_0)$. These values are shown as green vertical lines in Figure \ref{fig:merge5} (e-h) and Figure \ref{fig:merge6} in section \ref{sec:additional_plots}. The estimates from TreeBUGS have remarkably small variance. Although there are observable variations among methods, {the estimated surprises $\hat {p}_2(Y_0^{(n)})$ are all consistently less than 0.05. This means $Y_0^{(n)}$ seems surprising under $M_2$, indicating} that all methods unanimously support the PD model ($M_1$). This indicates that according to the estimated Bayes factors, once controlled processing fails, our stereotype (prejudice) is activated before resorting to a random guess.

\vspace{-1cm}

\section{Conclusion}\label{sec:conclusion}
In this work, we proposed a Deep Bayes Factor estimator in the absence of explicit likelihood. We established two types of consistency for Bayes factor estimators: estimation and inferential.  In this paper, we focused on comparing two models. However, extending DeepBF to multiple models is feasible. For multiple competing models, a multi-class classifier can be trained to yield the posterior model probabilities \(\{\pi(M_j\C Y^{(n)}_0)\}_{j=1}^K\), which also can inform the best model for $Y^{(n)}_0$. Additionally, to compute Bayes factors between any two models, \(M_i\) and \(M_k\), the ratio \(\pi(Y^{(n)}_0\C M_i)/\pi(Y^{(n)}_0\C M_k)\) can be derived easily using \(\{\pi(M_j\C Y^{(n)}_0),\pi(M_j)\}_{j=1}^K\). 

Now we mention a {few} future directions for this work. {In this study, we did not focus on optimizing learning hyperparameters, including the network architecture. Our goal was not to fine-tune these aspects to peak performance. Hence, we kept  hyperparameters such as the learning rate or the network designs the same across all datasets and settings. Future work could benefit from advanced tuning and   more sophisticated network designs. Indeed, }
we think further exploration of neural network architecture is desirable. The current feed-forward networks do not guarantee order invariance, while Bayes factors are order invariant for i.i.d. models. Using set-invariant networks, like {DeepSets} in \cite {zaheer2017deep} may provide a reasonable solution. As $n$ increases, the distribution of the true Bayes factors becomes simpler, while the model complexity concurrently rises. Addressing this paradox may involve attention-based set invariant networks   \citep{lee2019set} where  the size of feature maps does not increase with the input dimension. {A preliminary investigation of this idea using DeepSets is presented in Section \ref{sec:deepset} of the Supplement, which demonstrates promising results.}

{While there is room for enhancing the scalability of DeepBF,   our work is immune to the conventional requirement of large datasets for deep learning training. Our approach decouples the number of observations $n$ from the training sample size $T$. The training sample size $T$ is specified by the user based on their  time and computational budget. On the other hand, our framework can be applied even with a single observation $n=1$. We believe our work not only extends the applicability of Bayes factors  but also extends the applicability of deep learning within Bayes. }



\setlength{\bibsep}{0pt plus 0.3ex}
\bibliographystyle{abbrvnat}
\bibliography{ref}

\newpage

\appendix
\setcounter{page}{1}
\setcounter{equation}{1}
\renewcommand{\thesection}{S\arabic{section}}
\renewcommand{\thefigure}{S\arabic{figure}}
\renewcommand{\thetable}{S\arabic{table}}
\renewcommand{\theequation}{S\arabic{equation}}

\begin{center}
    {\Large \bf Supplementary Material}
\end{center}

\begin{abstract}
    This supplementary material contains proofs of all the theorems in the main text as well as the description of the rank-based ABC algorithm (in Section \ref{sec:ABC_}). In particular, the proof of the consistency of our Deep Bayes Factor estimator in Theorem \ref{cor:consistency} is presented in Section \ref{pf:consistency}. The proofs for its consistency as a model selection tool in Theorem \ref{thm:convP} is presented in Section \ref{pf:convP}. Practical considerations of choosing the deep learning architecture are elaborated in Section \ref{sec:comp}. {Section \ref{sec:deepset} demonstrates promising results achieved by using DeepSets for Deep Bayes Factors. Section \ref{sec:exp_detail_section} provides experimental details,} and Section \ref{sec:other_datasets} provides some additional empirical investigations, and Section \ref{sec:additional_plots} contains some additional visualizations and tables. 
\end{abstract}
\section{Proofs of Theorems}

\subsection{Proof of Lemma \ref{thm:consistency_BF}}\label{sec:proof_lemma}
Below we reformulate the assumptions from \cite{kaji2022metropolis} in the context of {\em marginal likelihood ratio} estimation.

Recall that  for two discriminators $D_1$ and $D_2$, we define a Hellinger-type discrepancy 
$$
d_H(D_1,D_2)=\sqrt{h(D_1,D_2)^2+h(1-D_1,1-D_2)},
$$
where 
$h(D_1,D_2)=\sqrt{(E_1+E_2)(\sqrt{D_1}-\sqrt{D_2})^2}$ and where $E_1$ and $E_2$ are expectations under the marginal distributions
under model $M_1$ and $M_2$, respectively.

We also define a ``Hellinger-style" neighborhood of the oracle classifier $D^*$ within an approximating sieve $\mathcal D_n$ 
defined in \eqref{eq:optimal_D} as (for some $\delta>0$)
$$
\mathcal D_{n}^*(\delta)=\{D\in \mathcal D_n: d_H(D,D^*)\leq \delta\}.
$$
The following assumption requires that the training classification algorithm can find a sufficiently good approximate maximizer.
At the same time, it requires that the approximating sieve is not too complex.
\begin{ass}\label{ass:kaji1}
Assume that there exists an estimator $\hat D_T$ that satisfies
$$
\M_T(\hat D_T)\geq \M_T(D^*)-O_{\bar P}(\delta_{T,n}^2)
$$ 
for a non-negative sequence $\delta_{T,n}$ where $\M_T(\cdot)$ was defined in \eqref{eq:loss}.
Moreover, assume that the bracketing entropy integral (defined in Section 1 in Appendix of \cite{kaji2022metropolis}) satiesfies
$J_{[]}(\delta_{T,n},\mathcal D_{n}^*(\delta_{T,n}),d_H)\lesssim \delta_{T,n}^2\sqrt{T}$ and that there exists $\alpha<2$ such that
$J_{[]}(\delta,\mathcal D_{n}^*,d_H)/\delta^\alpha$ has a majorant decreasing in $\delta$. 
\end{ass}
Under this assumption \cite{kaji2020adversarial} conclude that 
$d_H(\hat D_T,D^*)=O_{\bar P}(\delta_{T,n})$, where $\bar P$ encompasses all randomness in the ABC reference table (simulated data from models $M_1$ and $M_2$).
 The next assumption is needed to show consistency of the Bayes factor estimator and can be regarded as a version of a bounded marginal likelihood condition.
 More details can be found in \cite{kaji2022metropolis}.

\begin{ass}\label{ass:kaji2}
There exists $M>0$ such that for $k=1,2$ we have  $E_k(P_1/P_2)^j\leq M$  for $j=1,2$ and
$$
\sup_{D\in\mathcal D_{n}^*(\delta_{T,n})}E_k\left(\frac{D^*}D\big| \frac{D^*}{D}\geq \frac{25}{16}\right)<M\quad\text{and}\quad
\sup_{D\in\mathcal D_{n}^*(\delta_{T,n})}E_k\left(\frac{1-D^*}{1-D}\big| \frac{1-D^*}{1-D}\geq \frac{25}{16}\right)<M
$$
for $\delta_{T,n}$ from Assumption \ref{ass:kaji1}.
\end{ass}
Going back to the proof of Lemma \ref{thm:consistency_BF}, we write 
\begin{align*}
E_1\left| \log \widehat{BF}_{1,2}(Y)-\log  {BF}_{1,2}(Y)\right|&\leq E_1\left| \log \frac{\hat D_T}{D^*}\right|+E_1\left|
\log\frac{1-D^*}{1-\hat D_T}\right|
\end{align*}
Following the same steps as in the proof of Theorem 4.1 in \cite{kaji2022metropolis}
with notation $P_0=E_1$ and $D_\theta=D^*$ it can be verified\footnote{Their proof presents a variant of the results with the expectation inside the absolute value.  Inspecting the proof, 
however, it follows that these statements hold with the expectation outside the absolute value as well. } that for any $D\in \mathcal D_{n}^*(\delta_{T,n})$
$$
E_1\big |\log \frac{D}{D^*}\big |\leq 2(1\vee \sqrt{M})\delta_{T,n}\quad\text{and}\quad E_1 \big |\log\frac{1-D}{1-D^*}\big |\leq 
(\sqrt{2}+2+25M/\e)\sqrt{2M}\delta_{T,n}.
$$
Analogously, with $P_0=E_2$ we obtain the statement above for $E_2$.
Because under the Assumption
\ref{ass:kaji1}  we have  $d_H(\hat D_T,D^*)=O_{\bar P}(\delta_{T,n})$, we 
conclude that for $j=1,2$ we have $E_{j}\left| \log \widehat{BF}_{1,2}(Y)-\log  {BF}_{1,2}(Y)\right|=O_{\bar P}(\delta_{T,n})$.

 \subsection{Proof of Theorem \ref{cor:consistency}}\label{pf:consistency}
We focus on the case when $p^*(\cdot)=\pi_1(\cdot\C\theta_1^*)$ for some $\theta_1^*\in\R^{d_1}$ as the second case is analogous. 
For an event 
$$
\mathcal A_n=\{Y^{(n)}_0: \log   BF_{1,2}(Y_0^{(n)})\geq 1/ \nu_n \}
$$ 
we have $P^\infty_*[\mathcal A_n^c]=o(1)$ from  Definition \ref{def:consist}. On the event $\mathcal A_n$, using the notation
$$
u(Y_0^{(n)})\equiv \log \widehat{BF}_{1,2}(Y_0^{(n)})-\log  {BF}_{1,2}(Y_0^{(n)}),
$$ 
we can write
$$
\log \widehat{BF}_{1,2}(Y_0^{(n)}) \geq -|u(Y_0^{(n)})|+\log BF_{1,2}(Y_0^{(n)})\geq  -|u(Y_0^{(n)})|+1/  \nu_n.
$$
Therefore
\begin{equation}
  P^\infty_* \left[\log \widehat{BF}_{1,2}(Y_0^{(n)}) <1/(2\nu_n)   \right]\leq P^\infty_*\left[|u(Y^n_0)|> 1/(2\nu_n)  \right] +P^\infty_*[\mathcal A_n^c ].\label{eq:aux2}
\end{equation}
Next, for   $\Delta=1/(2\nu_n)>0$ we define $C_n(\Delta)=\{Y_0^{(n)}:|u(Y_0^{(n)})|>\Delta\}$  and  write
$$
P^\infty_*(|u(Y_0^{(n)})|>\Delta )=\int_{C_n(\Delta)}  \prod_{i=1}^np^*(Y_i^0) \d Y_0^{(n)}= \int_{C_n(\Delta)} \frac{\pi(Y_0^{(n)}\C M_1)}{r(Y_0^{(n)})}  \d Y_0^{(n)},
$$
where (adopting $K_n\equiv K_ n^{(1)}$ from Assumption \ref{ass:KL})
$$
 {r(Y_0^{(n)})}= {\int_{\R^{d_1}} \prod_{i=1}^n\frac{\pi_1(Y_i^0\C\theta_1)}{p^*(Y_i^0)}\pi_1(\theta_1)\d\theta_1}>  \int_{K_ n } \prod_{i=1}^n\frac{\pi_1(Y_i^0\C\theta_1)}{p^*(Y_i^0)}\pi_1(\theta_1)\d\theta_1.
$$
For any $\delta>0$, denote by $ D_n(\delta)=\{Y_0^{(n)}: n^{d_1/2}r(Y_0^{(n)})>1/ \delta \}$ then it follows from 
the proof of Theorem 4.1 in \cite{tokdar2010bayesian}  that  under Assumption \ref{ass:KL} in the main text
$$
P^\infty_*(D_n^c(\bar\delta_n))\leq P^\infty_*\left( n^{d_1/2} \int_{K_ n } \prod_{i=1}^n\frac{\pi_1(Y_i^0\C\theta_1)}{p^*(Y_i^0)}\pi_1(\theta_1)\d\theta_1\leq 1/{\bar \delta_n}\right)
\leq \frac{c}{(\log(C\bar\delta_n)-c)^2}=o(1)
$$ for any $\bar\delta_n\rightarrow\infty$.
Then  
$$
P^\infty_*\left(|u(Y_0^{(n)})|>\Delta\right)\leq n^{d_1/2}\bar\delta_n\int_{C_n(\Delta)\cap D_n(\bar\delta_n)}    {\pi(Y_0^{(n)}\C M_1)}  \d Y_0^{(n)}+ o(1).
$$
Next, from Lemma \ref{thm:consistency_BF} we have
\begin{align*}
n^{d_1/2}\bar\delta_n\int_{C_n(\Delta)\cap D_n(\delta_n)}    {\pi(Y\C M_1)}  \d Y&
\leq n^{d_1/2}\bar\delta_n P(|u(Y)|>\Delta\C M_1)\\
&\leq \frac{n^{d_1/2}  \bar\delta_n}{\Delta} E_1|u(Y)|=\frac{n^{d_1/2}\bar \delta_n}{\Delta}  \delta_{n,T} O_{\bar P}(1).
\end{align*}
Going back to \eqref{eq:aux2}, we find for $\Delta=1/(2\nu_n)$
$$
P^\infty_* (\log \widehat{BF}_{1,2}(Y_0^{(n)}) <1/(2\nu_n))\leq 2\nu_n {n^{d_1/2}\bar \delta_n}  \delta_{n,T} O_{\bar P}(1) +o(1).
$$
We obtain the desired statement under the assumption that $\nu_n{n^{d_1/2}\bar \delta_n}  \delta_{n,T}=o(1)$.
A similar expression holds for the second term. $\qedhere$

\subsection{Proof of Theorem \ref{thm:convP}}\label{pf:convP}

We can write
$
 \widehat{BF}_{1,2}(Y_0^{(n)})=  {BF}_{1,2}(Y_0^{(n)})\times \e^{u(Y_0^{(n)})},
$
where
\begin{equation}
u(Y_0^{(n)})=\log \frac{\hat D_{T}}{D^*} -\log \frac{1-\hat D_{T}}{1-D^* }= \log \widehat{BF}_{1,2}(Y^{(n)}_0)-\log  {BF}_{1,2}(Y^{(n)}_0)\label{eq:u}
\end{equation}
and where  with $\hat D_T$ defined in \eqref{eq:estimator} and $D^*$ in \eqref{eq:optimal_D}.
Next, we follow the steps of Theorem 4.1 in \cite{tokdar2010bayesian}.  We utilize the following  identity valid for any set $A_n^1\subset \R^{d_1}$ of parameters $\theta_1$ and $Y\in \mathcal Y^n$
$$
\pi(Y \C M_1)=\frac{\int_{A_n^1}\pi_1(Y \C\theta_1)\pi_1(\theta_1)\d\theta_1}{\pi(A_n^1\C Y ,M_1)}.
$$
We can thus write
$$
 \widehat{BF}_{1,2}(Y_0^{(n)})=\e^{u(Y_0^{(n)})}\times \frac{\int_{A_n^1}\pi_1(Y_0^{(n)}\C\theta_1)\pi_1(\theta_1)\d\theta_1}{\int \pi_2(Y_0^{(n)}\C\theta_2)\pi_2(\theta_2)\d\theta_2}\times \frac{1}{\pi(A_n^1\C Y_0^{(n)},M_1)}.
$$
  From Assumption \ref{ass:conc} we assume that there exists $\theta^*\in\R^{d_1}$ such that for
 $A_n^1=\{\theta_1\in\R^{d_1}: d(\theta_1,\theta^*)<\varepsilon_n\}$ we have
 $$
 -\log[\pi(A_n^1\C Y^{(n)}_0,M_1)]=o_{P^{\infty}_*}(1).
 $$
Let $K_n=K_n^{(2)}$ be as in Assumption \ref{ass:KL} and write
 $$
\log  \widehat{BF}_{1,2}(Y_0^{(n)})\leq u(Y_0^{(n)})+\log  \frac{\int_{A_n^1}\prod_{i=1}^n\frac{\pi_1(Y_i^0\C\theta_1)}{p^*(Y_i^0)}\pi_1(\theta_1)\d\theta_1}{\int_{K_n}\prod_{i=1}^n\frac{\pi_2(Y_i^0\C\theta_2)}{p^*(Y_i^0)}\pi_2(\theta_2)\d\theta_2}+o_{P^{\infty}_*}(1).
$$
Next,  under the model $M_2$ we can write $p^*(\cdot)=\pi_2(\cdot\C\theta^*_2)$ for some $\theta_2^*\in\R^{d_2}$.
Similarly as in the proof of  Theorem 4.1 in \cite{tokdar2010bayesian},   Assumption \ref{ass:KL} then implies that for any $\delta>0$
\begin{equation}
P^\infty_*\left(n^{d_2/2}\int_{K_n}\prod_{i=1}^n\frac{\pi_2(Y_i^0\C\theta_2)}{p^*(Y_i^0)}\pi_2(\theta_2)\d\theta_2\leq 1/\delta\right)\leq \frac{c}{(\log (C \delta)-c)^2}\label{eq:aux}
\end{equation}
which implies  
$$
P^\infty_*\left(-\log\left[{n^{d_2/2}\int_{K_n}\prod_{i=1}^n\frac{\pi_2(Y_i^0\C\theta_2)}{\pi_2(Y_i^0\C\theta^*_2)}\pi_2(\theta_2)\d\theta_2} \right]\geq \log(\delta)\right)\leq \frac{c}{(\log (C \delta)-c)^2},
$$
meaning that $-\log\left[{n^{d_2/2}\int_{K_n}\prod_{i=1}^n\frac{\pi_2(Y_i^0\C\theta_2)}{\pi_2(Y_i^0\C\theta^*_2)}\pi_2(\theta_2)\d\theta_2} \right]=O_{P^\infty_*}(1)$.
Next, for any $\Delta>0$ we have 
$$
P^\infty_* \left[  n^{d_2/2} \int_{A_n^1}\prod_{i=1}^n\frac{\pi_1(Y_i^0\C\theta_1)}{p^*(Y_i^0)}\pi_1(\theta_1)\d\theta_1 > \Delta\right]\leq 1/\Delta n^{d_2/2}\pi_1(A_n^1)
$$
and thereby from Assumption \ref{ass:conc}    we have  
$$
\log \e^{1/\eta_n}n^{d_2/2}  \int_{A_n^1}\prod_{i=1}^n\frac{\pi_1(Y_i^0\C\theta_1)}{p^*(Y_i^0)}\pi_1(\theta_1)\d\theta_1=O_{P^\infty_*}(1).
$$ 
Next, similarly as in the proof of Theorem \ref{cor:consistency} we have for   $\Delta>0$   
and any arbitrarily slowly increasing sequence $\bar\delta_n$
$$
P^{\infty}_*(|u(Y^{(n)}_0)|>\Delta)\leq \frac{n^{d_2/2}\bar \delta_n}{\Delta}  \delta_{n,T} O_{\bar P}(1)+ o(1).
$$
This implies $|u(Y^n_0)|=n^{d_2/2}\bar \delta_n \delta_{n,T} O_{\bar P\times P^*_{\infty}}(1)$ and thereby, as long as $n^{d_2/2}\bar \delta_n \delta_{n,T}-1/\eta_n\rightarrow-\infty$ we have $\widehat{BF}_{1,2}(Y_0^{(n)})\rightarrow 0$ in $ P^*_{\infty}$-probability.
{\spacingset{1.5}
\begin{figure}
    \centering
    \includegraphics[width=0.7\linewidth]{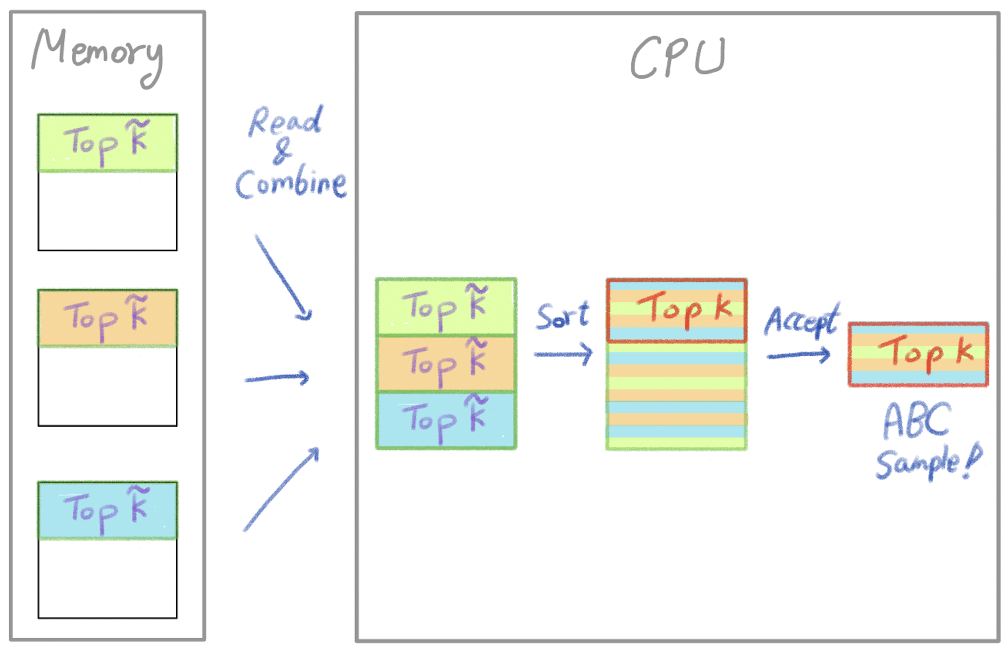}
    \caption{The stratified rank-based ABC algorithm to estimate Bayes factors in a computationally efficient way. It is not necessary that the distances for the entire $m$ samples are saved in the memory other than the the smallest $k$-distances and their labels. }
    \label{fig:stratified}
\end{figure}
}
\section{Rank-Based ABC algorithm}\label{sec:ABC_} 
Given a data sample $Y_0$, the widely used ABC algorithm for Bayes factor estimation for $BF_{1,2}(Y_0)$ accepts an ABC sample $\tilde Y$ if $\rho(\eta(Y_0),\eta(\tilde Y))\leq \epsilon$, where $\eta$ is a sufficient statistic, $\rho$ is a distance measure, and $\epsilon$ is the threshold. Choosing the right $\epsilon$ is a particularly challenging task. A too large $\epsilon$ reduces the quality of the estimator, while a too small $\epsilon$ increases computational burdens. To accept a pre-fixed number of samples, many more simulated samples will be required. Therefore, to step side the difficulty of tuning $\epsilon$, we use a rank-based acceptance rule. We first simulate $M$ samples $A = \{\tilde Y_1,...,\tilde Y_M\}$, and then compute the distances for all $Y_i$ as $\rho(\eta(Y_0),\eta(\tilde Y_i))$. Lastly, we take the samples of the $k$-smallest distances. Other than this new acceptance rule, this rank-based ABC estimator is identical to the ABC-based estimator explained in \cite{robert2011lack}.

Now, we discuss how to scale up the computation when estimating Bayes factors for multiple datasets. Note that the described rank-based computation can be done simultaneously. When our interest is to estimate Bayes factors for multiple data samples, e.g., $B=\{Y_1,...,Y_R\}$ for some integer $R$, the same $A$ can be used for ABC sampling to estimate $BF_{1,2}(Y_r)$ for any of $Y_r\in B$. When computing all distances at once through a tensor data structure, we can do all computation simultaneously. When $R$ is large, however, the size of the tensor generated by this calcultation can be computationally prohibitive. Therefore, we use a stratified version of the described algorithm. Instead of generating $M$ samples, we generate and save $m$ datasets, each of which consists of $\lfloor M/m\rfloor$ samples for some $m$ such that $M=m\times \lfloor M/m\rfloor$. Now, for each small strata dataset, we can compute the distances from $Y_0$ to each point in the dataset to know which points correspond to the smallest $\tilde{k}$ distances. We gather only the labels and the distances that correspond to the smallest $\tilde{k}$ distances. 
Once the calculations for each strata dataset are complete, we read in the data from memory and then combine those labels and distances at one centralized location. This consolidated data, which consists of the k smallest distances and their corresponding labels from each strata, should now be compact enough to manage within CPU memory. Then, among the gathered distances we can again find the smallest $k$ distances and their corresponding labels to compute the BF estimates in \eqref{eq:abc1}. Such estimates are a stratified version of the described rank-based ABC algorithm. The key idea of this stratified approach is illustrated in Figure \ref{fig:stratified}. The choice of those hyperparameters in our numerical results is in Table \ref{tab:ABC_spec}.

\begin{table}
    \centering
    \begin{tabular}{c||c|c|c|c}\hline
          Setting&$M$ (Total ABC samples)&  $m$ (number of strata)&  $\tilde k$& $k$\\\hline
          Figure \ref{fig:two_dim_good} &1,000,000&  1000&  5& 100\\\hline
          All other simulations&1,200,000&  1200&  5& 120\\\hline
    \end{tabular}
    \caption{ABC configurations for all our numerical examples. Because the top $k$ ABC samples are finally chosen, the estimation range is approximately $[1/k, k]$.}
    \label{tab:ABC_spec}
\end{table}

\section{Practical Considerations}\label{sec:comp}

Throughout the paper, we employed a fully connected feed-forward neural network (FNN), renowned for its universal approximability \citep{hornik1991approximation}. An FNN consists of a linear layer followed by a non-linear activation, which in our case is ReLU. We observed that expanding intermediate layers proportionally with input size can improve estimation but also introduce optimization instability. To address this, we utilized batch normalization neural networks (BNNs) \citep{ioffe2015batch}. In BNNs, each linear layer is followed by batch normalization before applying ReLU, except for the output layer, which uses a sigmoid activation. The complete design of our neural network is detailed in Section \ref{section:network}. For FNNs, the architecture includes an input layer mapping to a hidden space of size 64, followed by two \(64 \times 64\) linear layers with ReLU activations, and a final linear layer that maps to a single output with a sigmoid activation. Conversely, for BNNs, we upscale the first hidden layer based on the input size to improve estimation. Note that BNNs might show inconsistencies between training and testing due to different mean and variance handling. Therefore, during evaluation, we use normalization computed from a single mini-batch. We trained our models using the Adam optimizer \citep{kingma2014adam} with default hyperparameters (\(\beta_1 = 0.9\), \(\beta_2 = 0.999\)), an initial learning rate $0.01$, and a learning rate decay of 0.99 every 1000 iterations. The training loss function did not include regularization, and the networks were trained without dropout for $M=400,000$ iterations using a mini-batch size of 200, resulting in a total of $T=80,000,000$ samples.


\subsection{Network architecture}\label{section:network}
{\spacingset{1.2}
\begin{lstlisting}
import torch 
from torch import nn

class FNN(nn.Module):
    def __init__(self,input_shape,factor=64, depth = 2):
        super(FNN,self).__init__()
        self.fc1 = nn.Linear(input_shape,factor)
        self.linears = nn.ModuleList([nn.Linear(factor,factor) 
        			for i in range(depth)])
        self.fc4 = nn.Linear(factor,1)
        
    def forward(self,x):
        x = torch.relu(self.fc1(x))
        for md in self.linears:
            x = torch.relu(md(x))
        x = torch.sigmoid(self.fc4(x))
        return x

    def BF_est(self,x,eps = 0):
        d = self.forward(x)
        return (d+eps)/(1+eps-d))
        
class BNN(nn.Module):
    def __init__(self, input_shape, factor=64, depth=2):
        super(BNN, self).__init__()
        self.fc1 = nn.Linear(input_shape, factor)
        self.bn1 = nn.BatchNorm1d(factor)
        current_factor = factor
        self.linears = nn.ModuleList()
        self.bns = nn.ModuleList()

        for _ in range(depth):
            next_factor = max(int(current_factor * reduction_ratio), 16)
            self.linears.append(nn.Linear(current_factor, next_factor))
            self.bns.append(nn.BatchNorm1d(next_factor))
            current_factor = next_factor
        
        self.fc4 = nn.Linear(current_factor, 1)
        
    def forward(self, x):
        x = self.fc1(x)
        x = torch.relu(self.bn1(x))
        
        for md, bn in zip(self.linears, self.bns):
            x = md(x)
            x = torch.relu(bn(x))
            
        x = torch.sigmoid(self.fc4(x))
        return x
 
    
\end{lstlisting}
       }
\section{DeepSet for Deep Bayes Factors}\label{sec:deepset}
{ In this section, we apply the DeepSet network in \cite{zaheer2017deep} as a $q$-dimensional feature extractor before feeding it into an FNN. In this case, the input dimension for the FNN is $q$ (set as $q=2$ in our implementation). The DeepSet and FNN together compose the discriminator denoted by $D^{(t)}$ in Algorithm \ref{alg:original}. The structure of the DeepSet is shown at the end of this section (Section \ref{section:network_deep}). The benefit of this structure is that the DeepSet feature extractor makes the entire $D^{(t)}$ order-invariant among observations, and the size of the network does not scale up as $n$ increases.

For the setting in Example \ref{ex:toy}, we draw a figure corresponding to Figure \ref{fig:two_dim_good} and Figure \ref{fig:two_dim_good2} with the automatic sufficient statistics learning using DeepSet, for increasing sufficient statistics dimensions in Figure \ref{fig:two_dim_good_deep} (high probability area) and Figure \ref{fig:two_dim_good2_deep} (wider range). As we can see, within the high probability area, the estimation is reasonable, closely matching the quality of FNNs for $n=2$.
\begin{figure}[!t]
    \centering
    \includegraphics[width=0.8\linewidth]{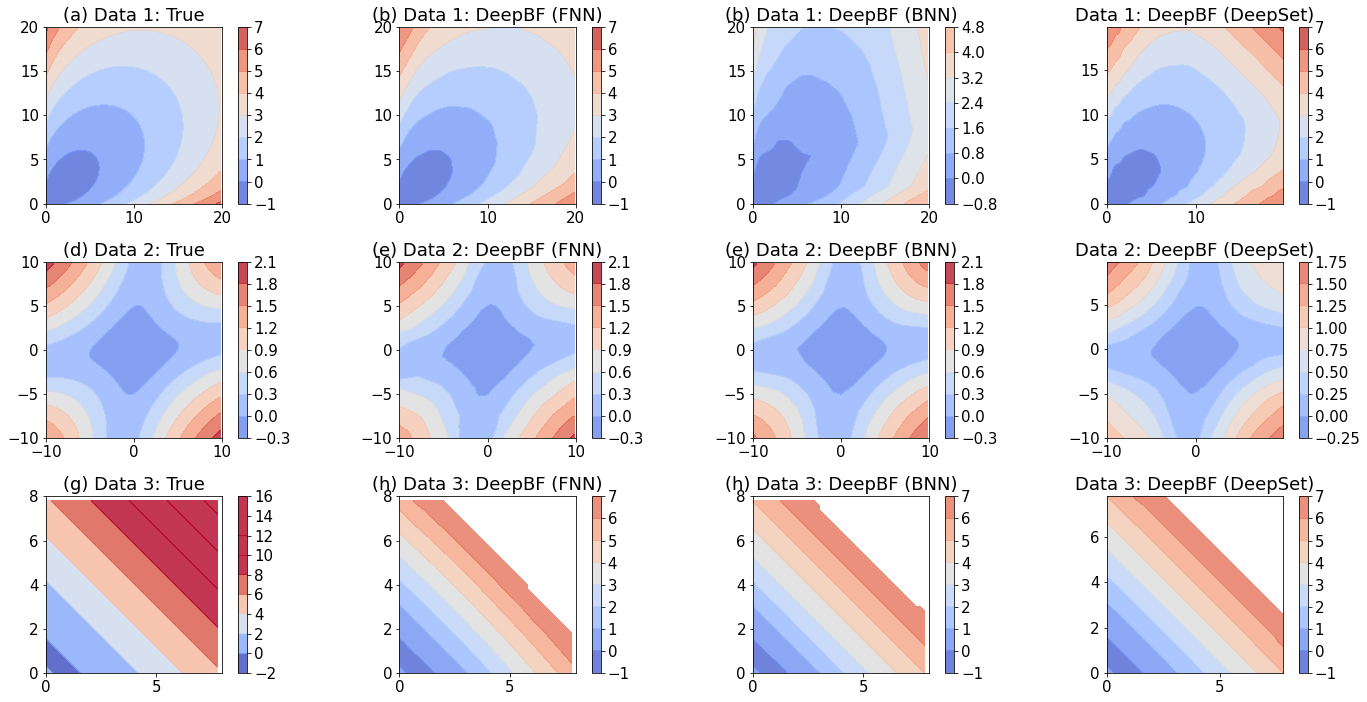}
    \caption{True and estimated {BF's} on the $\log_{10}$ scale ({DeepBF} with DeepSet) evaluated on a fine  grid of values $Y=(Y_1,Y_2)'$.  First row: Negative Binomial vs Poisson distributions. Second row: {Gaussian vs Gaussian mixture priors}. Third row: Hierarchical exponential distribution.}
    \label{fig:two_dim_good_deep}
\end{figure}

\begin{figure}[!h]
    \centering
    \includegraphics[width=.8\linewidth]{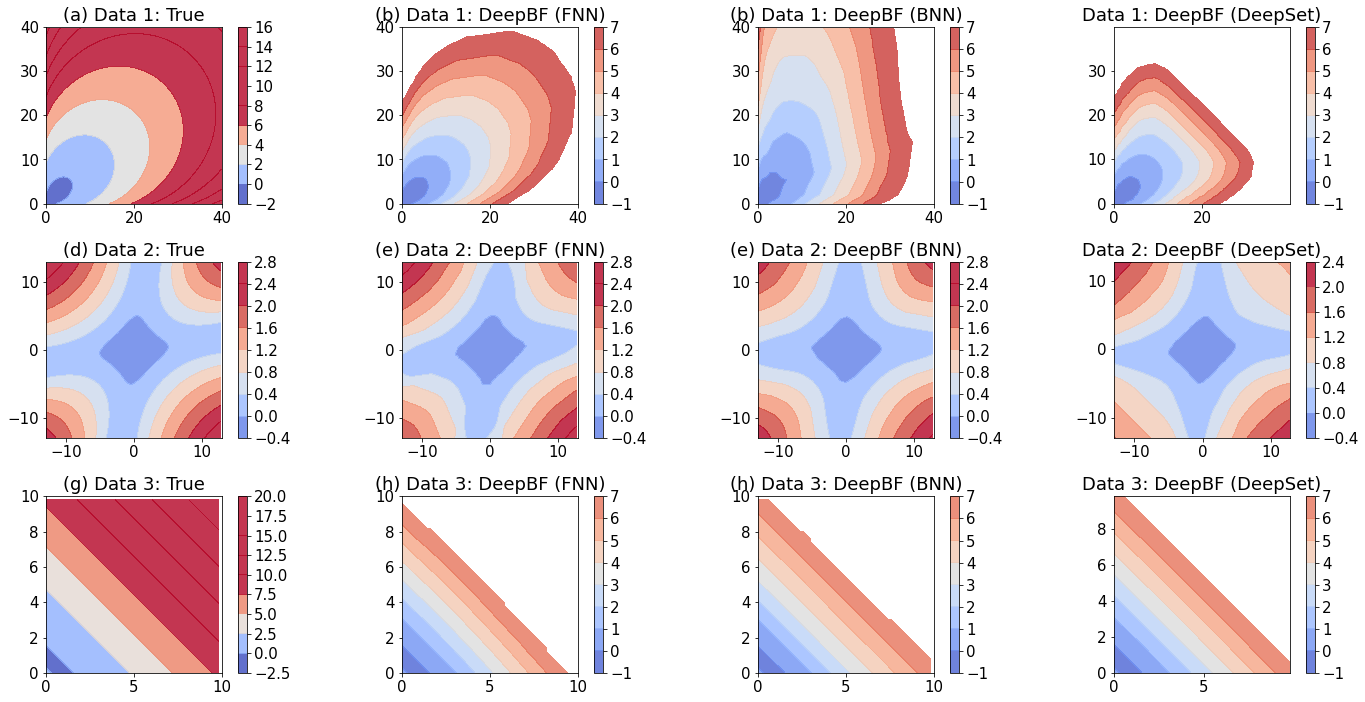}
    \caption{The scatter plot of random 30,000 samples on the visualization in Figure \ref{fig:two_dim_good_deep}.}
    \label{fig:two_dim_good2_deep}
\end{figure}

We also apply the DeepSet design to the settings considered in Section \ref{sec:simul} (Dataset 1 (i) and Dataset 1 (ii)) and Section \ref{sec:other_datasets} (Dataset 3). We observed that the discriminator tends to collapse to a constant function as $n$ increases. This issue can be mitigated by using a larger DeepSet network architecture, which can be achieved by stacking multiple DeepSet layers, as described in Section \ref{section:network_deep}. In this simulation, we stacked up two DeepSet layers. To further prevent collapse, we reduced the training mini-batch size to 300. Various alternative approaches to stabilize learning could also be considered, such as gradient clipping. To know the success of learning, we can use the classification loss in \eqref{eq:loss} or classification accuracy as a proxy. This allows us to select the best model from multiple runs with different random initializations.

The results of this approach are presented in Figure \ref{fig:final_deepset}. The reported model was selected based on its highest classification accuracy on a test dataset (size $3000$) among 10 random runs. The benefits of using DeepSet for DeepBF extend beyond simply satisfying order invariance. This design enables the DeepBF estimator to handle much larger values of $n$. While we observe a decline in estimation quality as $n$ increases, the inferential consistency remains intact. Our theoretical results and proofs suggest that increasing the capacity of the DeepSet network—enabling it to represent an ideal classifier—could further enhance estimation quality. However, this would require longer training durations and the development of more stable learning techniques and hyperparameter configurations to reliably train larger networks.

\begin{figure}[!h]
    \centering
    \includegraphics[width=.9\linewidth]{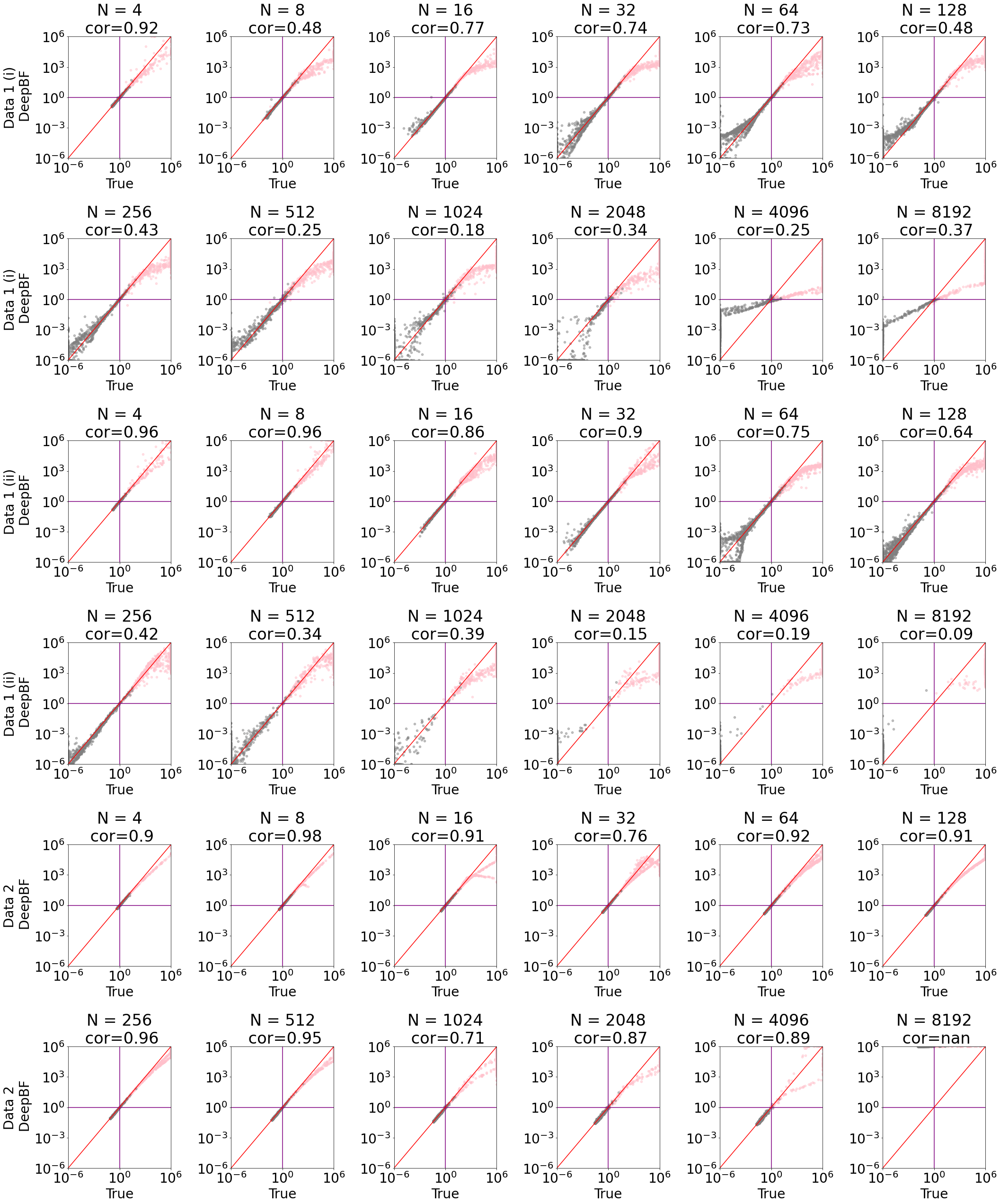}
    \caption{The scatterplot of random 3,000 samples by DeepBF with the DeepSet architecture. It scales to much larger $n$ than the FNNs and BNNs (both computational time and learnability). In the plot, ``cor" indicates the correlation coefficient between the true and estimated Bayes factor values.}
    \label{fig:final_deepset}
\end{figure}

\subsection{DeepSet Network architecture}\label{section:network_deep}
The DeepSet architecture for DeepBF is ``DeepSet2" which uses two ``basic\_DeepSet" layers. On the output of ``DeepSet2", a Net (FNN) is applied to get the final one dimensional output.
{\spacingset{1.2}
\begin{lstlisting}
import torch 
from torch import nn

class basic_DeepSet(nn.Module):
    def __init__(self, input_dim, ss_dim, 
                  c_factor = 64,c_n_layers=3, 
                 device="cuda"):
        super(basic_DeepSet,self).__init__()
        self.common_feature_net = MLP(device=device,
                                      dim=input_dim,
                                      z_dim = ss_dim,
                                      dropout=0,
                                      factor=c_factor,
                                      n_layers=c_n_layers) 
        self.to(device)
        self.device = device
        
    def forward(self, x):
        shape = x.shape
        assert len(shape)==3
        phi = self.common_feature_net(x.view(-1,shape[-1])).view(x.shape[0],x.shape[1],-1).mean(1)
        return phi
    
class DeepSet2(Net):
    def __init__(self, N_y, ss_dim=2, factor=64, depth=2, ss_dim_med=16, device="cuda"):

        super(DeepSet2, self).__init__(ss_dim, factor, depth)
        
        self.suff_net1 = basic_DeepSet(device=device,
                                       input_dim = 1,
                                       ss_dim = ss_dim_med)
        
        self.suff_net2 = basic_DeepSet(device=device,
                                       input_dim = 1,
                                       ss_dim = ss_dim)
        
    def forward(self,x):
        med = self.suff_net1(x).unsqueeze(-1)
        med2= self.suff_net2(med)
        return super().forward(med2) 
    
class Net(nn.Module):
    def __init__(self,input_shape,factor=64, depth = 2):
        super(Net,self).__init__()
        self.fc1 = nn.Linear(input_shape,factor)
        self.linears = nn.ModuleList([nn.Linear(factor,factor) for i in range(depth)])
        self.fc4 = nn.Linear(factor,1)
    def forward(self,x):
        x = torch.relu(self.fc1(x))
        for md in self.linears:
            x = torch.relu(md(x))
        x = torch.sigmoid(self.fc4(x))
        return x
\end{lstlisting}
}
}

\section{Examples}\label{sec:exp_detail_section}
\subsection{Details of Example \ref{ex:toy}}\label{example_details}
All of these examples assume an iid data setup. Data 1 explores classical Bayesian modeling with Negative Binomial and Poisson distributions \citep{BERN}. Define \( M_1 \)  through \( Y_i|p \sim NB(1, p) \), where \( p \sim Beta(\alpha_1,\beta_1) \), and \( M_2 \) with \( Y_i|\lambda \sim Pois(\lambda) \) where \( \lambda \sim Gamma(\alpha_2, \beta_2) \). We set \( \alpha_1 = \beta_1 = \alpha_2 = \beta_2 = 1 \). In this case, \cite{robert2011lack} identified   bias in ABC Bayes factors \eqref{eq:abc1} when a summary statistic is used. {Data 2 considers {Gaussian mixture data with a Gaussian prior on the means}. {Namely, \( M_1: Y_i \sim  N(\mu_{11},2^2)/2+ N(\mu_{12},2^2) /2 \) with \( \mu_{11} \sim N(2, 1.5^2) ,~\mu_{12}\sim \frac{1}{2}N(-2, 1.5^2) \), and \( M_2: Y_i \sim N(\mu_2, 2.5^2) \) with \( \mu_2 \sim N(0, 1) \).} Data 3 presents a nested model test, distinguishing between a hierarchical Bayesian model  \( M_1: Y_i|\lambda \sim Exp(\lambda) \) with \( \lambda \sim Gamma(2,2) \)  and a fixed-parameter model   \( M_2: Y_i \sim Exp(3) \). We can conceptualize \( M_2 \) as a Bayesian hierarchical model with a Dirac prior \(\delta(\lambda-3)\).}  To nicely visualize heatmaps of the results, we start with two observations, i.e. \(n=2\), and present results for larger $n$ later in Section \ref{sec:simul}. The heatmaps zoom in 
onto values $Y=(Y_1,Y_2)'$ that are most likely under either one of the two considered models (See Figure \ref{fig:two_dim_good2} in Section \ref{sec:additional_plots} for a broader range of $Y$). The ABC implementation is detailed in Section \ref{sec:ABC_}. In the effort to see the best ABC performance, we intentionally use the full data $Y$ as a summary statistic. {For the deep neural network architecture, we use an FNN.} Details on the implementations and deep learning architectures are in Section \ref{sec:comp}. 

\subsection{Toy Example for Model Criticism}\label{toy_continue_3}

{\spacingset{1.5}
\begin{figure}
    \centering
    \includegraphics[width=0.8\linewidth]{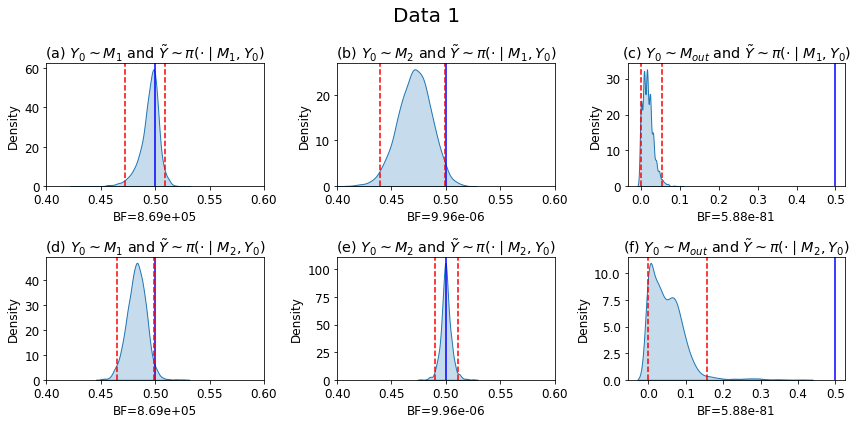}
    \vspace{-.5cm}
    \caption{Model Criticism on Data 1. The empirical density (KDE) plot of $Z(\tilde Y,\hat d)$ values defined in \eqref{eq:Z_dist}.  Red lines indicate the $2.5\%$ and $97.5\%$ quantiles of the simulated  $Z(\tilde Y,\hat d)$ values. The blue line represents $0.5$. }
  \label{fig:validation_ex}
\end{figure}
}

\begin{example}[Toy Example]\label{ex:toy3} Using data setups from Example \ref{ex:toy} and \ref{ex:toy_cont}, we train Bayesian GAN posterior generators \citep{wang2022adversarial}. 
For each dataset,  we  generate one fixed $Y_0^{(n)}\sim M_j$  and  then simulate $\tilde Y$ according to  $\pi(\cdot \C Y_0^{(n)}, M_k)$
for   four model combinations  $(j,k)\in \{1,2\}\times \{1,2\}$ and  $n=128$. For $M_2$ of Data 3, we instead sample $\tilde Y\sim \pi(\cdot\C M_2)$ independently from $Y_0$.\footnote{The model parameter is fixed as opposed to random with a prior.}  Additionally, we manually set "outliers" as values $Y_0^{(n)}$  unlikely under the given model\footnote{  All outliers are in a form of $Y_{\rm out} = (\alpha,\alpha, \cdots, \alpha)\in \mathcal{Y}^n$ for a manually chosen $\alpha$. Dataset 1: $\alpha=18$, Dataset 2: $\alpha=10$, Dataset 3: $\alpha=2.5$.}. {In Figure \ref{fig:two_dim_good}, the first two dimensions of the outliers (red) and typical inliers (yellow) are marked as stars.} To simulate $Z(\tilde Y,\hat d)$ values defined in \eqref{eq:Z_dist}, we train a simple logistic regression classifier, where the {non-linear predictor}  is a quadratic function of a univariate input. {The classifiers are optimized for every new draw of $\tilde Y$ and evaluated on a new draw of $\tilde Y$ to compute the draw $Z(\tilde Y,\hat d)$.
{We had $3\,000$ draws for each model.}} 

The kernel density smoothed empirical distribution of $Z(\tilde Y,\hat d)$ is displayed in Figure \ref{fig:validation_ex} for Data 1, and in Figure \ref{fig:validation_ex2} and \ref{fig:validation_ex3} (Section \ref{sec:additional_plots}) for Data 2 and 3, respectively.  
The consonant cases   $\tilde Y\sim \pi (\cdot \C Y_0^{(n)},M_k)$ and $Y^{(n)}_0\sim \pi(\cdot\C M_k)$ for $k=1,2$ are illustrated on the diagonal of the Figure \ref{fig:validation_ex} (panels (a) and (e)).
 We can see that when the posterior predictive distribution matches  the true generating distribution of $Y_0^{(n)}$, the $Z(\tilde Y,\hat d)$ values are  concentrated near $0.5$. On the other hand, panels (b) and (d) show the dissonant cases where  the true generating model and predictive model are mismatched. In these plots,  {the $Z(\tilde Y,\hat d)$ values are more spread out and the value $0.5$ is barely included within the the $2.5\%$ and $97.5\%$ sample quantiles}. This indicates poor or borderline evidence for the considered model suggesting further model comparisons through Bayes factors. On the other hand, when $Y_0^{(n)}$ are implausible under the considered model, critiquing  the suggested model becomes  easier. This is evident from panels  (c) and (f) where  the kernel density plots of sampled values $Z(\tilde Y,\hat d)$ are far away from $0.5$.  
 Such a scenario  yields a more decisive rejection of the proposed model without the need for further Bayes factor model comparisons. It is important to note, however,   Bayes factor values evaluated at  observed data $Y^{(n)}_0$ implausible under both models  may still strongly prefer one model over the other, despite the fact that predicted values from both modes are not compatible with $Y^{(n)}_0$. Therefore, it would seem that   Bayes factor model comparisons could be enhanced with our Z-value approach which  offers a careful preliminary examination of model adequacy before proceeding to model selection based on Bayes factors.
\end{example}

\section{Results on Additional Datasets}\label{sec:other_datasets}
{In this section, we present the results on the exponential-Gamma test dataset (Data 3) in Example \ref{ex:toy} and \ref{ex:toy_cont}. We do not consider the Gaussian vs Gaussian mixture (Data 2), as the computation of true Bayes factor is involved as $n$ increases.} We compare our methods with the ABC method described in Section \ref{sec:ABC_}. The scatter and KDE plots of our trained DeepBFs are presented in Figure \ref{fig:other78}. In the scatter plot, we see many points are aligned along the red reference line of perfect estimation. Here, for extremely large true BF values of Data 3, our method tend to overestimate. However, the KDE plot implies these points are extreme outliers because the KDE plots of the estimates are closely resemble the KDE plots of the true BFs. That is, for the vast majority of the datasets, the estimates are precise. In Figure \ref{fig:other78_3} we also present the Spearman's rho correlation (rank-based), the empirical KL divergence between the true and estimated distributions, and the MSE of log BFs. Also, in Figure \ref{fig:other78_2}, the ROC curves for $n=2$ and $n=128$ are presented together with the surprise MSE measure. Both figures demonstrate that our DeepBF is promising, outperforming ABC-based estimates. 

\spacingset{1.5}
\begin{figure}
    \centering
    \includegraphics[width=\linewidth]{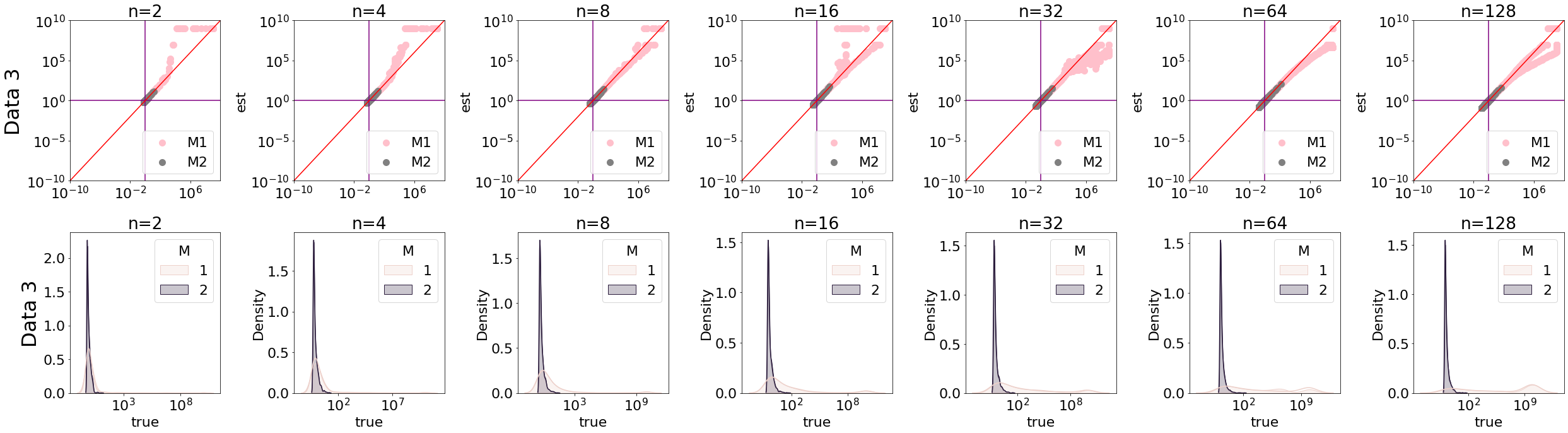}
    \caption{The scatter and KDE plots of DeepBF-FNN on the exponential-Gamma test dataset in Example \ref{ex:toy} and \ref{ex:toy_cont}. Thresholded at $10^{10}$.}
    \label{fig:other78}
\end{figure}

\begin{figure}
    \centering
    \includegraphics[width=\linewidth]{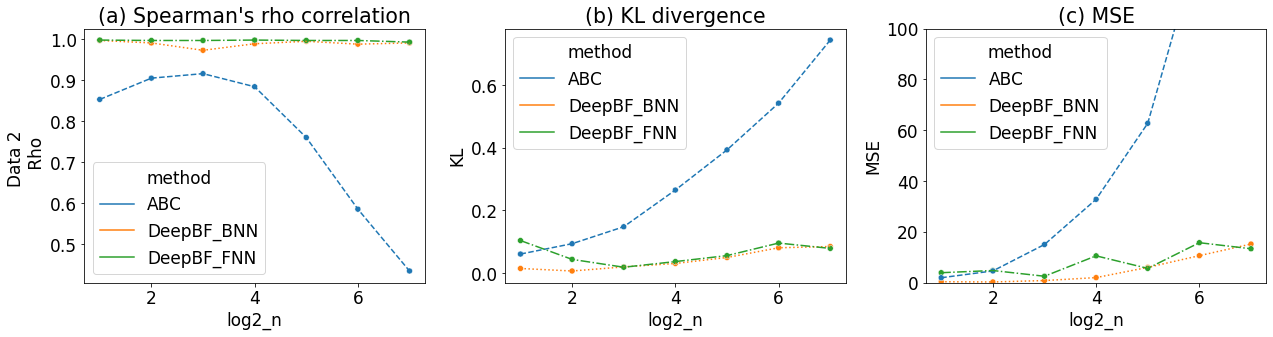}
    \caption{Evaluation of BF approximations for Data 3 (Exponential vs Gamma test datasets) in Example \ref{ex:toy} and \ref{ex:toy_cont}.}
    \label{fig:other78_3}
\end{figure}

\begin{figure}
    \centering
    \includegraphics[width=\linewidth]{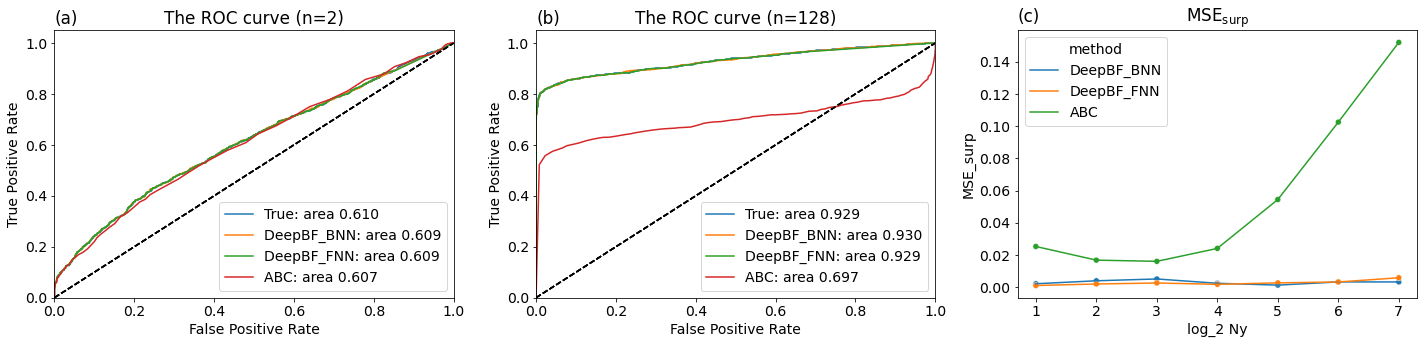}
    \caption{For Data 3. (a-b) ROC curves from true and estimated Bayes factors; (b) MSE for surprise measures.}
    \label{fig:other78_2}
\end{figure}

\begin{figure}
    \centering
    \includegraphics[width=.4\linewidth]{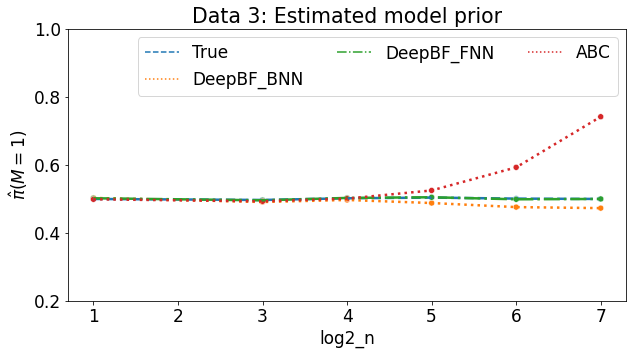}
    \caption{The estimated prior on Data 3 (Exponential vs Gamma test datasets) in Example \ref{ex:toy} and \ref{ex:toy_cont}. Closer to 0.5 is better.}
    \label{fig:other78_4}
\end{figure}

\newpage
\pagebreak

\section{Additional plots and tables}\label{sec:additional_plots}
Here, we show additional plots and tables not shown in the main body.

\begin{figure}[!h]
    \centering
    \includegraphics[width=.8\linewidth]{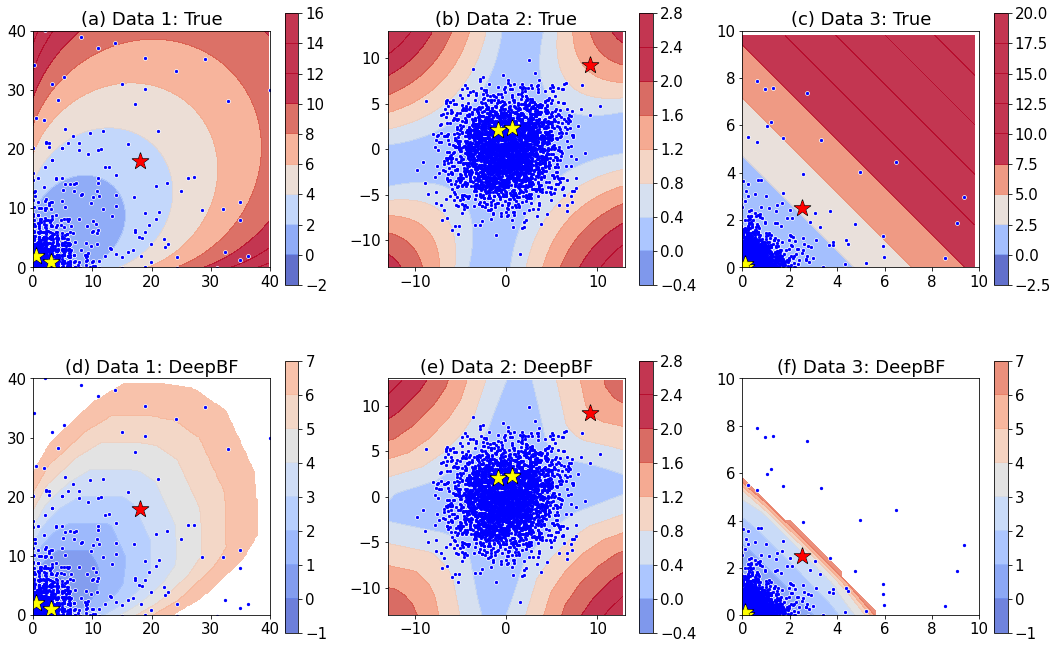}
    \caption{The scatter plot of random 30,000 samples on the visualization in Figure \ref{fig:two_dim_good}. For Data 1, slight noise is added to visualize as the data is discrete. The white areas represent machine infinity or NaN due to numerical overflow. These are the areas where datasets are unlikely to be realized and the BF estimates may behave less stably. We can see that  a large portion of the data  support for both models is in the areas where BF are numerically stable.  Only a small fraction of the samples falls outside the displayed domain: 4.42\% for Data 1, 0.08\% for Data 2, and 0.37\% for Data 3.}
    \label{fig:two_dim_good2}
\end{figure}

\begin{figure}[!h]
    \centering
    
    \includegraphics[width=\linewidth]{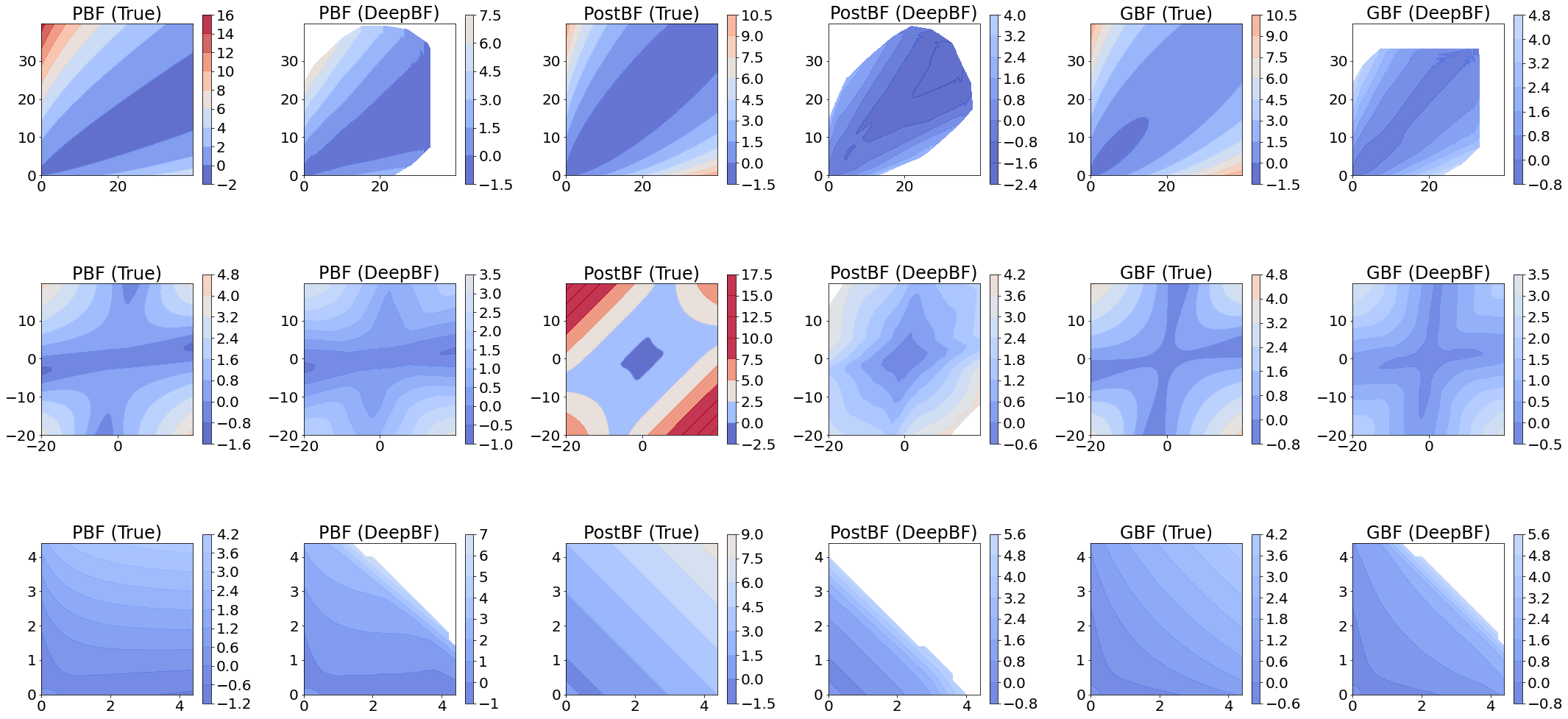}
    \caption{Visualization corresponding to Figure \ref{fig:partial1} on   a wider range of the input space (values on the $\log_{10}$ scale). The white areas represent machine infinity or NaN due to numerical overflow. These are the areas where datasets are unlikely to be realized and the BF estimates may behave less stably.}
    \label{fig:robust}
\end{figure}

\begin{figure}[!t]
    \centering
    \includegraphics[width=.9\linewidth]{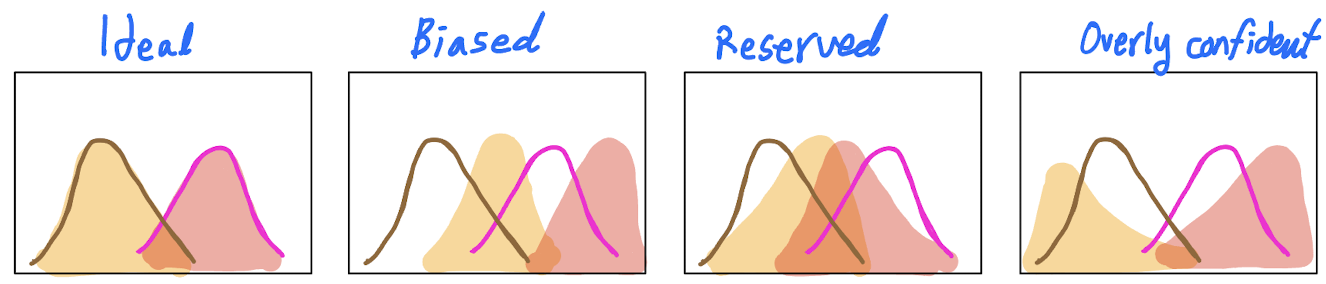}
    \vspace{-.5cm}
    \caption{Thick line: the true BF distributions on   samples simulated from $M_1$ or $M_2$ (depending on the color); Colored area: the estimated BF distributions on samples simulated from $M_1$ or $M_2$ (depending on the color).}
    \label{fig:four_types}
\end{figure}

\begin{figure}[!h]
    \centering
    \includegraphics[width=.7\linewidth]{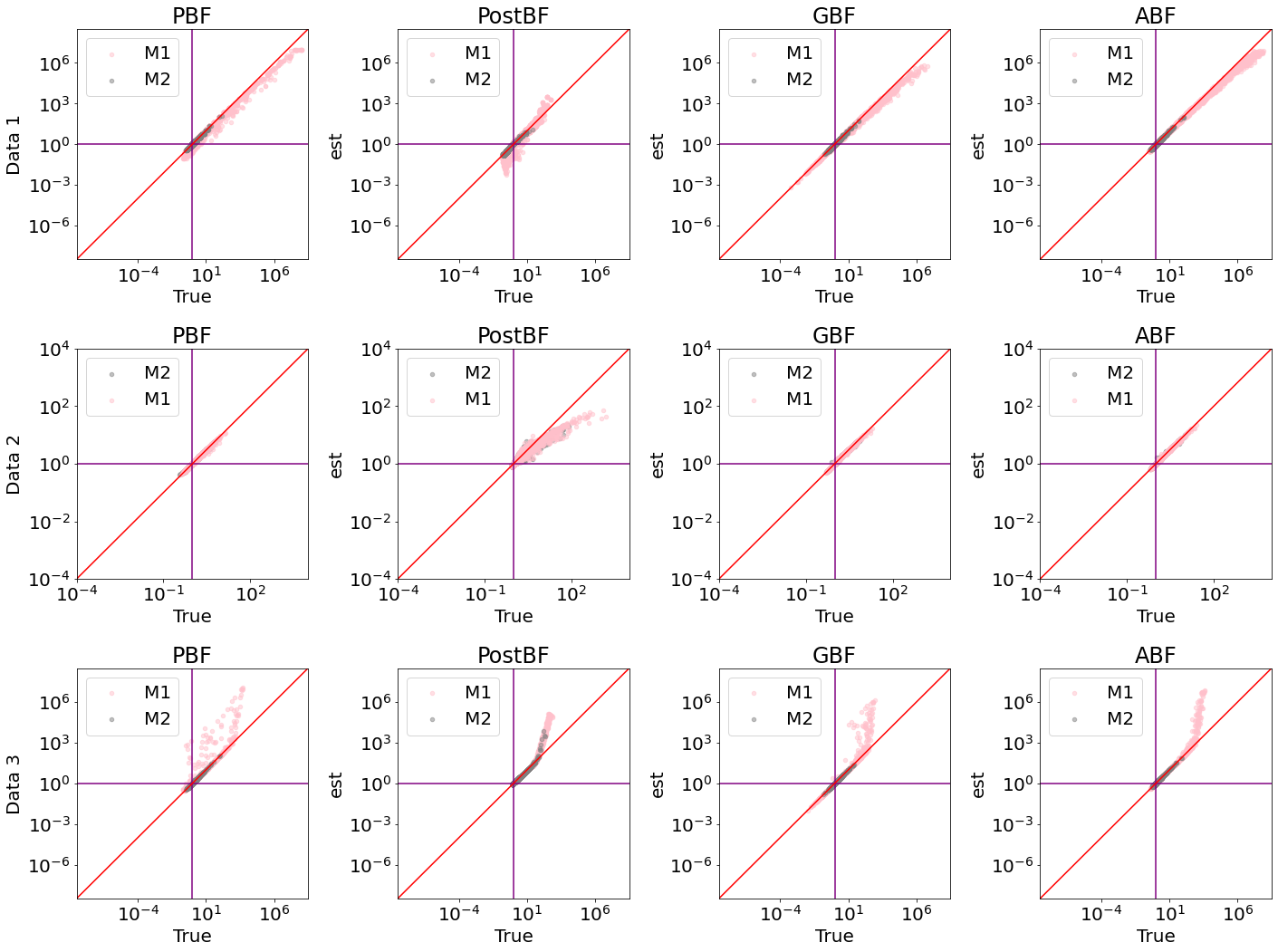}
    \caption{The scatter plot of Bayes factor variants for the three datasets in Example 
    \ref{ex:toy_cont}. }
    \label{fig:partial3}
\end{figure}

\begin{figure}[!h]
    \centering
    \includegraphics[width=.9\linewidth]{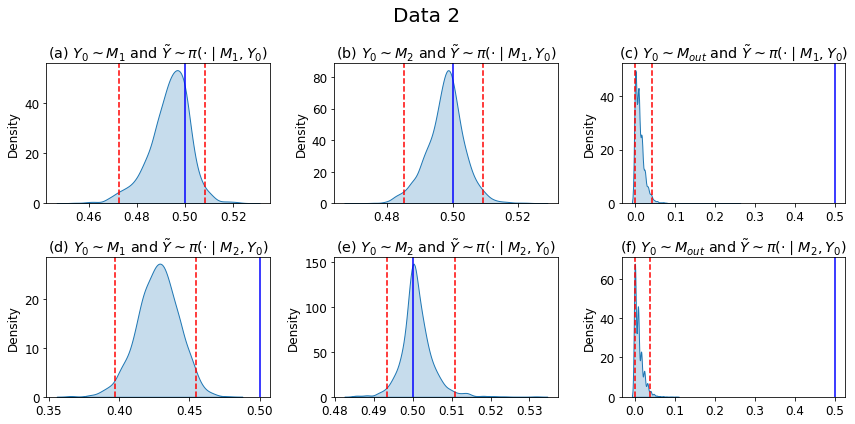}
    \caption{Example Data 2. The empirical density (KDE) plot of $Z(\tilde Y,\hat d)$ values defined in \eqref{eq:Z_dist}.  Red lines indicate the $2.5\%$ and $97.5\%$ quantiles of the simulated  $Z(\tilde Y,\hat d)$ values. The blue line represents $0.5$.}
    \label{fig:validation_ex2}
\end{figure}

\begin{figure}[!h]
    \centering
    \includegraphics[width=.9\linewidth]{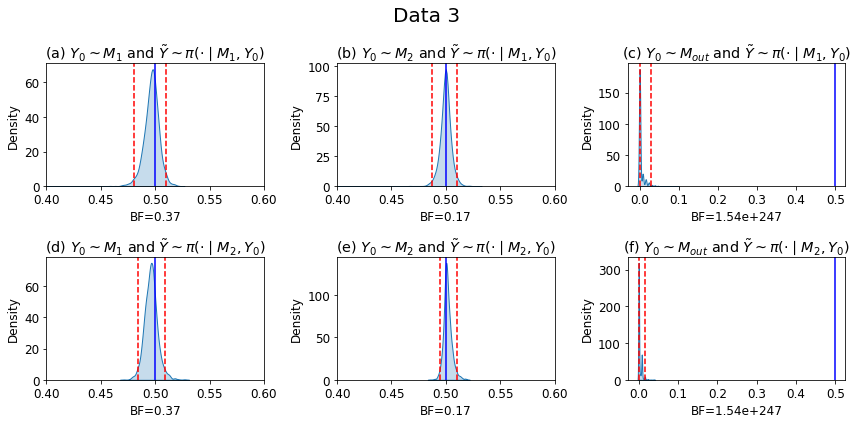}
    \caption{Example Data 3. The empirical density (KDE) plot of $Z(\tilde Y,\hat d)$ values defined in \eqref{eq:Z_dist}.   Red lines indicate the $2.5\%$ and $97.5\%$ quantiles of the simulated  $Z(\tilde Y,\hat d)$ values. The blue line represents $0.5$.}
    \label{fig:validation_ex3}
\end{figure}

\begin{figure}[!h]
    \centering
    \includegraphics[width=.7\linewidth]{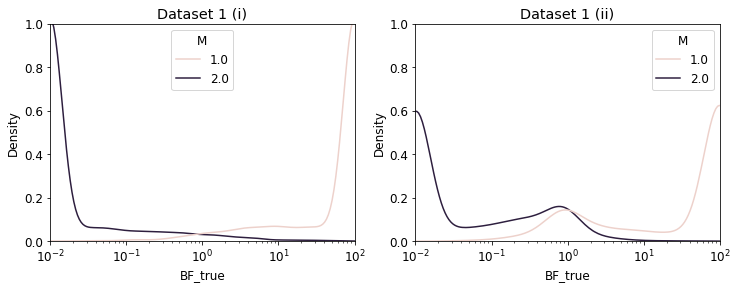}
    \caption{The KDE plot of the true Bayes factors for Dataset 1 (i) and (ii). The values are trimmed at 0.01 and 100. Dataset 1 (i) has a lower density near Bayes factor 1 compared with Dataset 2 (ii).}
    \label{fig:separability}
\end{figure}

\begin{figure}[!h]
    \centering
    \includegraphics[width=\linewidth]{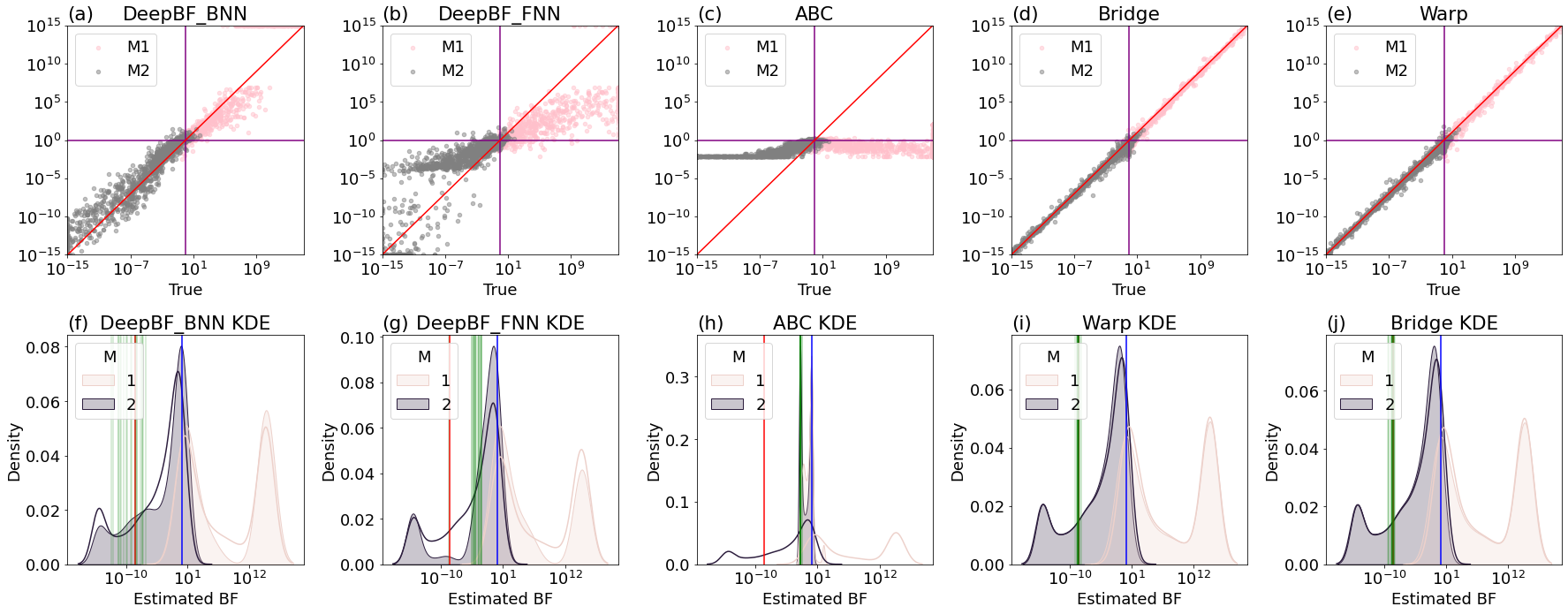}
    \caption{Data 1 (ii). Example of Negative Binomial vs. Poisson distributions when $n=128$. First row: Scatter plots of estimated vs. true BFs. Second row: KDE plots on estimated BFs (colored regions) and on true BFs (lines). A red line denotes \( BF_{1,2}(Y_0) \) with $Y_0$ from $M_2$, green lines denote \( \widehat{BF}_{1,2}(Y_0) \) over 20 trials, and the blue line is at the decision threshold of 1. x-axis: Log-scale, with values thresholded to the range \( [10^{-15},10^{15}] \).}
    \label{fig:consis2}
\end{figure}

\begin{figure}[!t]
    \centering
    \includegraphics[width=\linewidth]{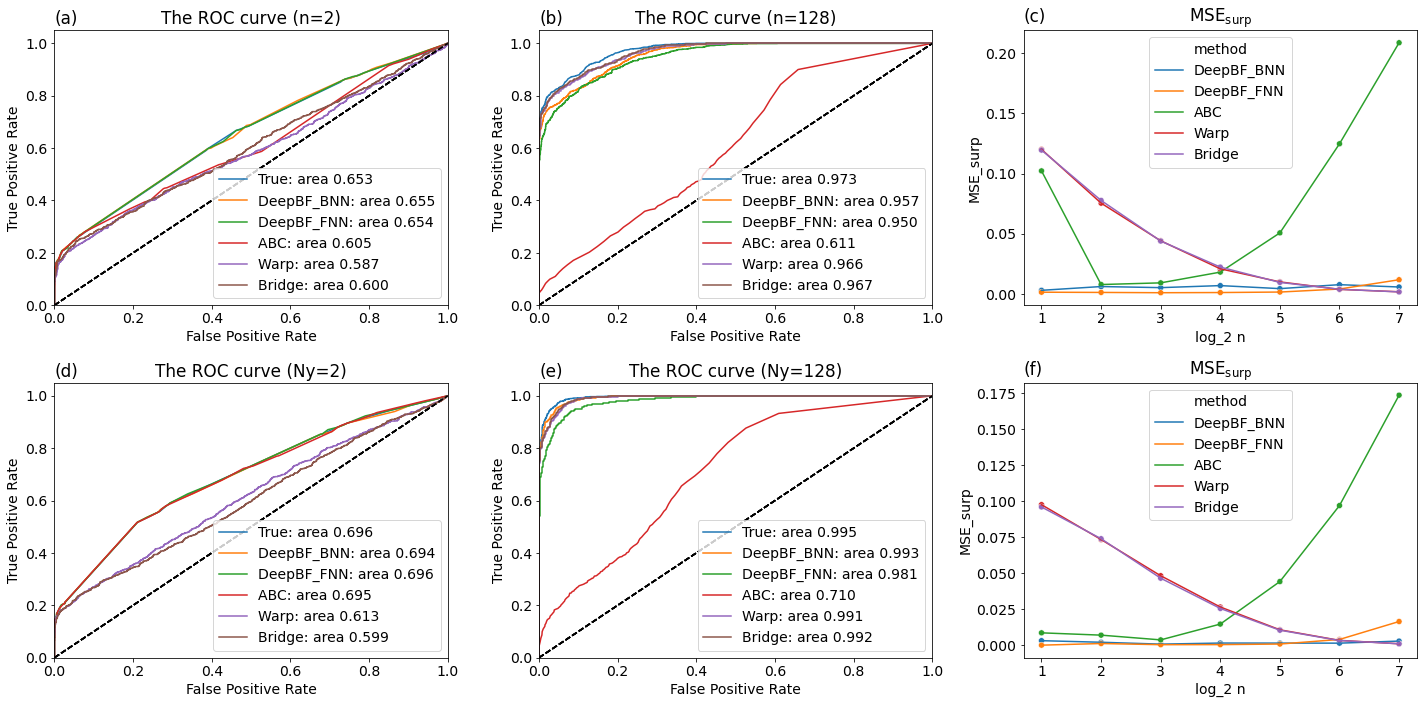}
    \caption{For Data 1 (i): (a-b) ROC curves from true and estimated Bayes factors; (c) MSE for surprise measures. Analogous results for Data 1 (ii) in (d-f).}
    \label{fig:roc}
\end{figure}

\begin{figure}[!h]
    \centering
    \includegraphics[width=\linewidth]{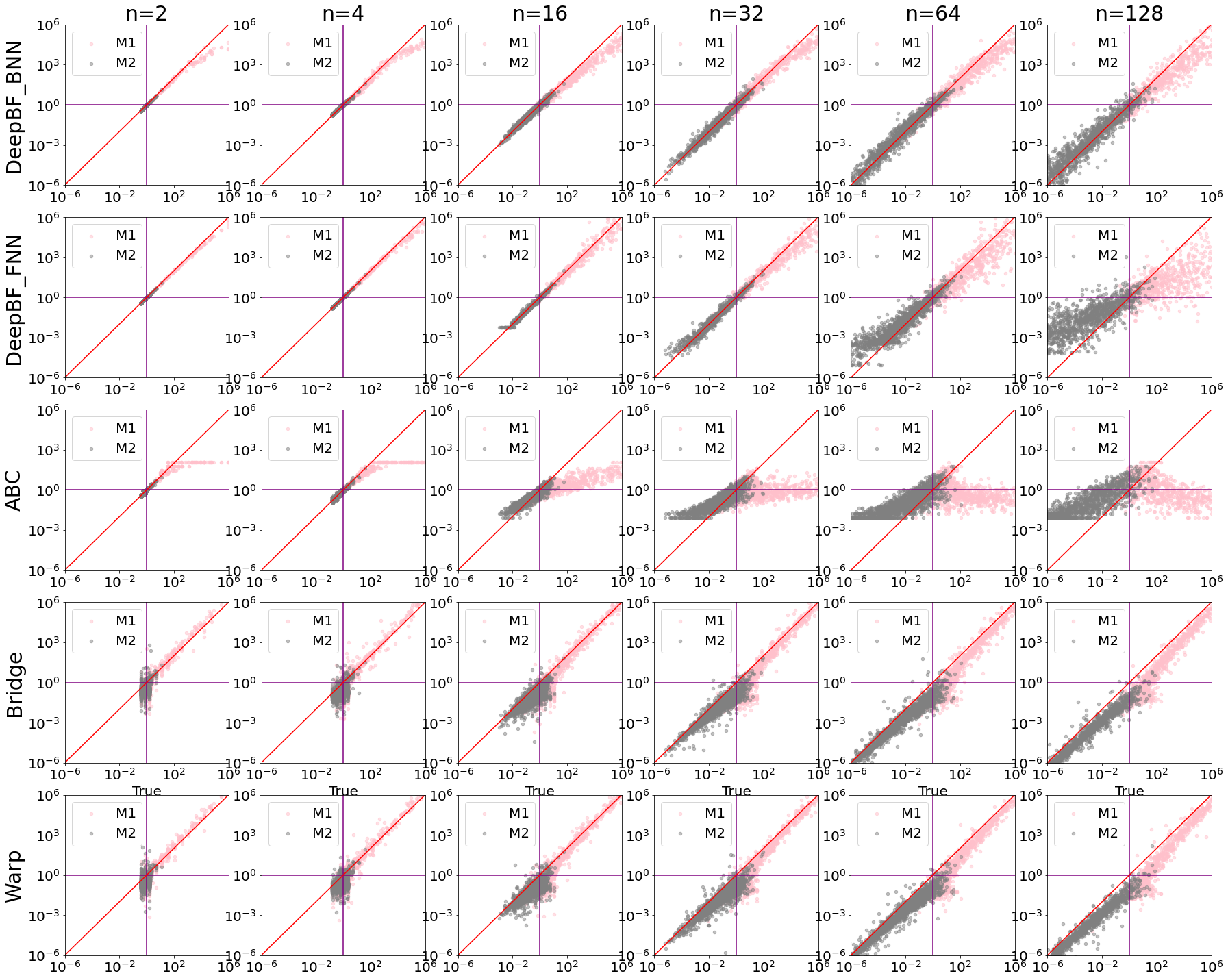}
    \caption{Data 1 (i). The scatter plot of the estimated BFs. The x and y axis: in log-scale with threshold at $\{10^{-6},10^6\}$.}
    \label{fig:scat2}
\end{figure}

\begin{figure}[!h]
    \centering
    \includegraphics[width=\linewidth]{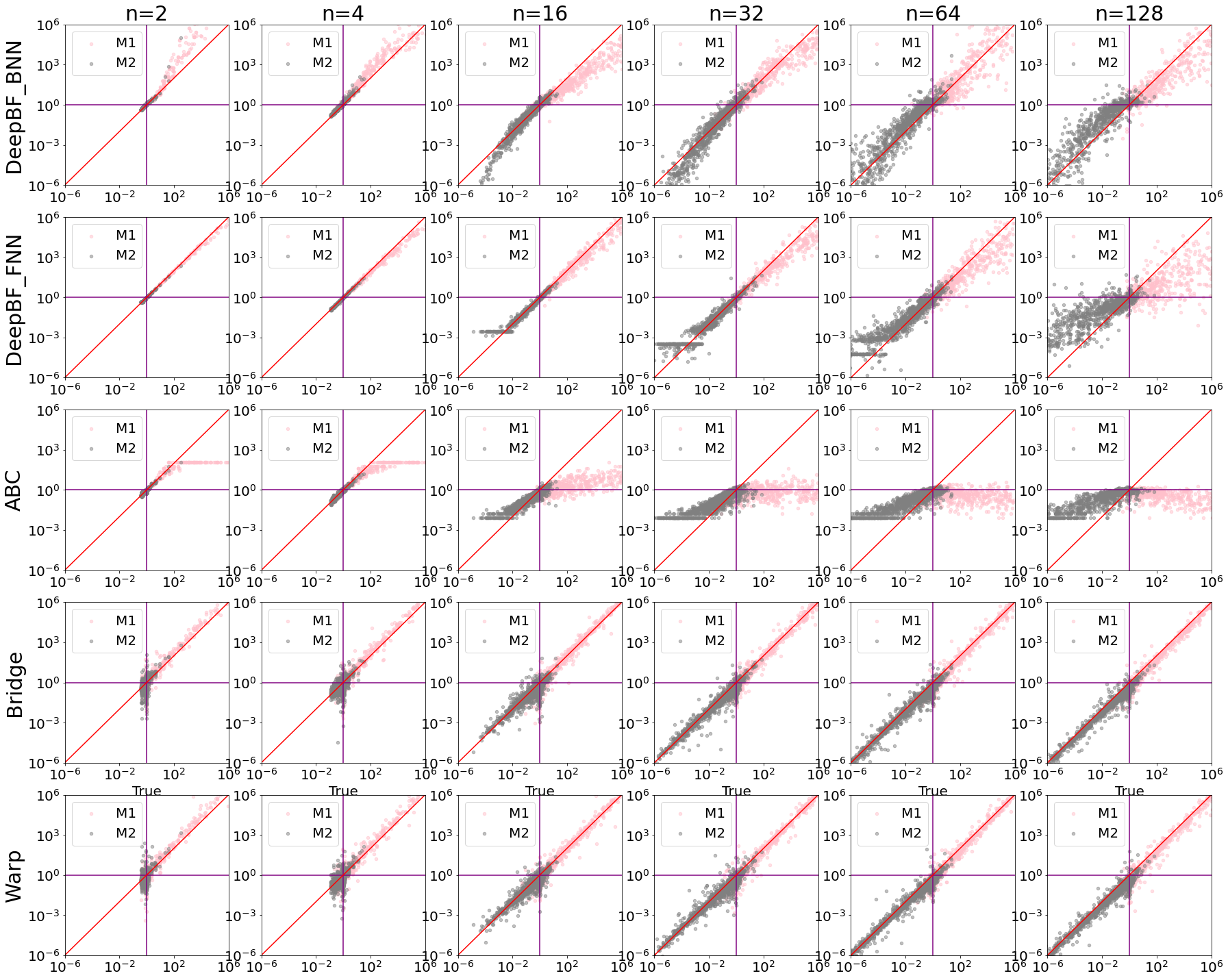}
    \caption{Data 1 (ii). The scatter plot of the estimated BFs. The x and y axis: in log-scale with threshold at $\{10^{-6},10^6\}$.}
    \label{fig:scat3}
\end{figure}

\begin{figure}[!h]
    \centering
    \includegraphics[width=\linewidth]{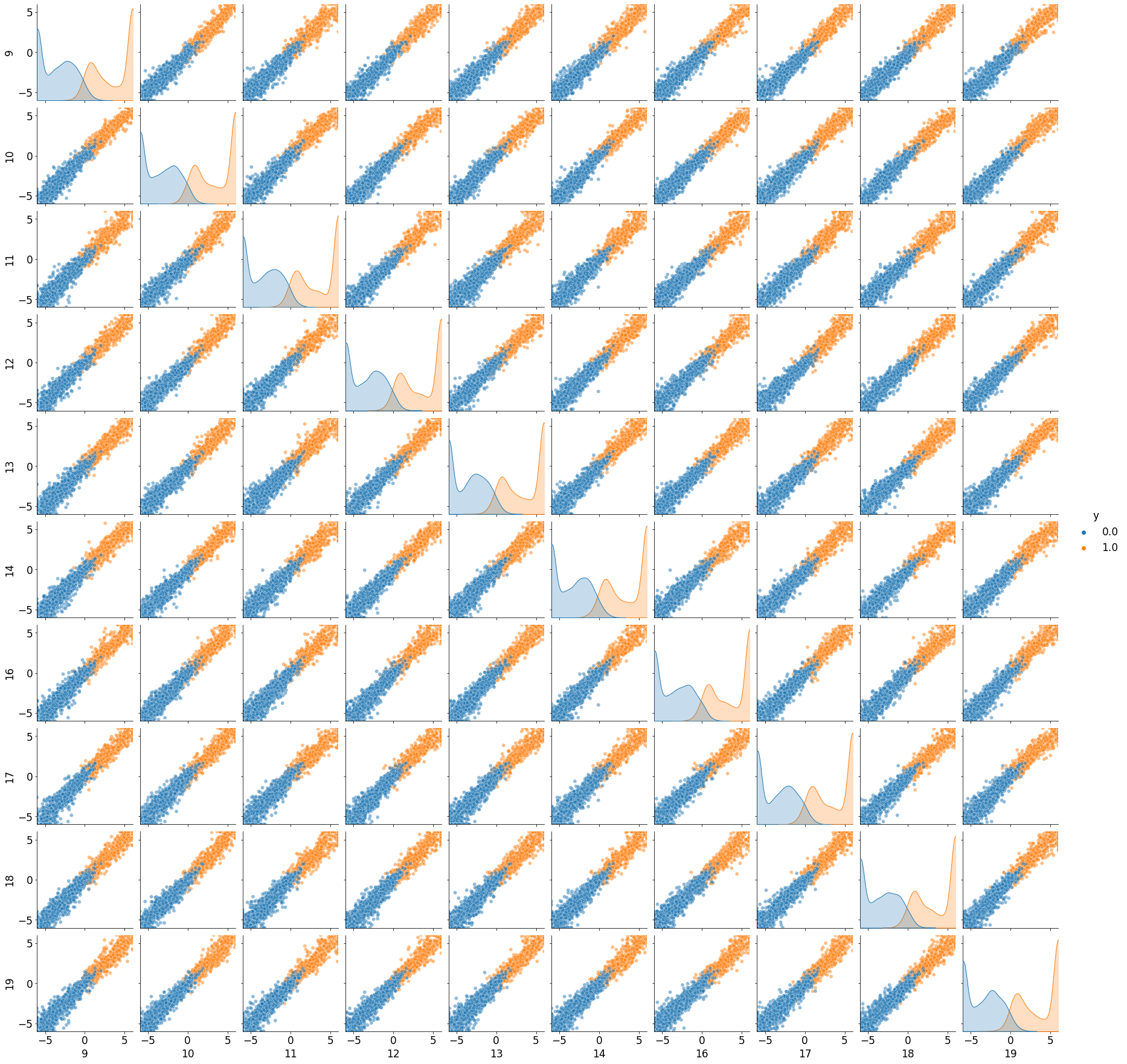}
    \caption{Dataset 1 (i). The scatter plots among multiple DeepBF-BNN estimates with $n=128$. Orange: data generated from $M_1$. Blue: data generated from $M_2$. }
    \label{fig:bnn-consist}
\end{figure}

\begin{table}
\centering\caption{The estimated model posteriors}
\begin{center}
\begin{tabular}{c || c | c  | c } 
 \hline
 Combination Type & Index (j) &$p_{j}$ in PDs& $p_{j}$ in Stroop\\
 \hline
 White Tool	&1	&C+(1-C)(A+(1-A)B)	&A+(1-A)(C+(1-C)B)\\
	White Gun	&2	&C+(1-C)((1-A)(1-B)	&(1-A)(C+(1-C)(1-B))\\
	Black Tool	&3	&C+(1-C)(1-A)(B)	&(1-A)(C+(1-C)B)\\
	Black Gun	&4	&C+(1-C)(A+(1-A)(1-B))	&A+(1-A)(C+(1-C)(1-B))\\
	Neutral Tool&	5	&C+(1-C)B	&C+(1-C)B\\
	Neutral Gun&	6&	C+(1-C)(1-B)&	C+(1-C)(1-B)\\
 \hline
\end{tabular}
\end{center}
\label{tb:pij}
\end{table}

\begin{figure}[!h]
    \centering
    \includegraphics[width=\linewidth]{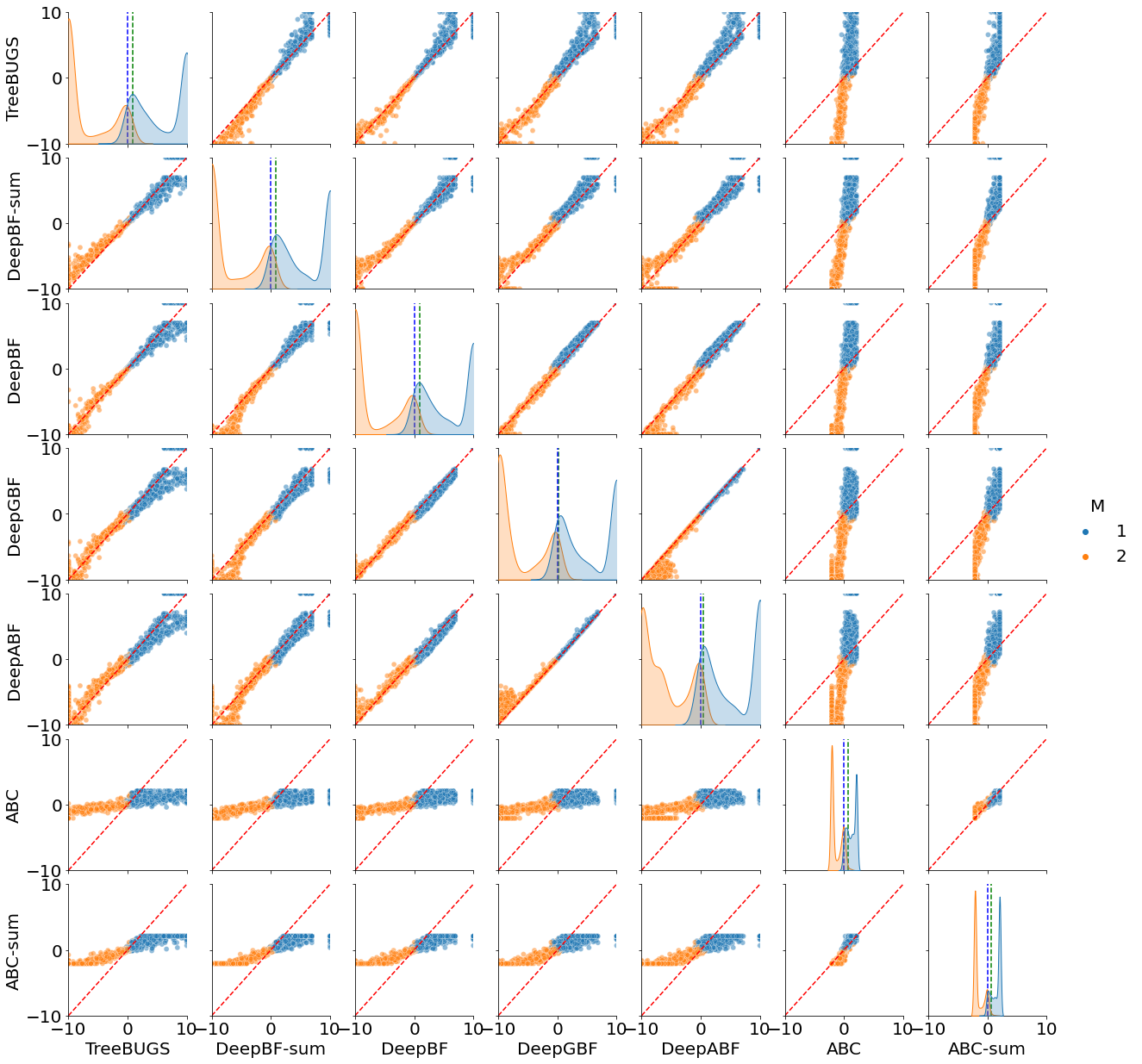}
    \caption{The scatter plot of the estimated BFs on 3,000 simulated data points among ABC, DeepBF, DeepABF, DeepBF, and TreeBUGS. Orange: data generated from $M_1$. Blue: data generated from $M_2$.}
    \label{fig:weapon_scatter}
\end{figure}

\begin{figure}[!h]
    \centering
    \includegraphics[width=0.8\linewidth]{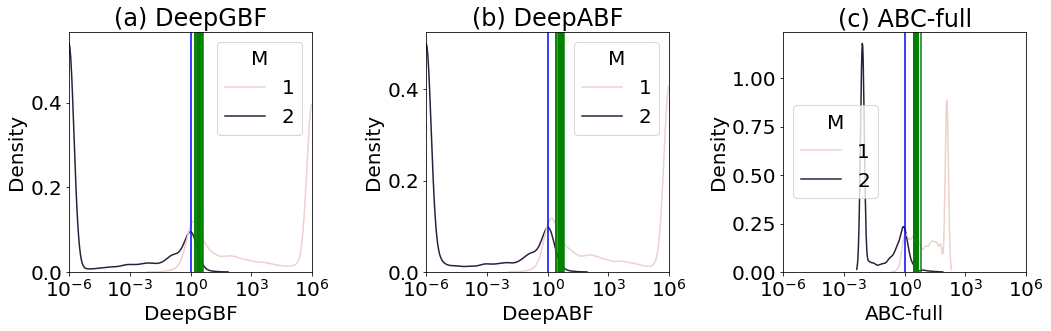}
    \caption{The empirical density (KDE) plot of the BF estimates with reference line at 1 (blue) and at the estimated $\widehat{BF}_{1,2}(Y_0)$ (green). }
    \label{fig:merge6}
\end{figure}

\end{document}